\documentclass[aps,prd,preprint,superscriptaddress,tightenlines,nofootinbib]{revtex4}

\usepackage{graphicx}% Include figure files
\usepackage{epsfig}% Include figure files
\usepackage{dcolumn}% Align table columns on decimal point
\usepackage{amsmath}% \text{} in math mode

\newcommand{\dstlnu}{D^*\ell\bar{\nu}}
\newcommand{\bdstlnu}{\bar{B}\to D^*\ell\bar{\nu}}
\newcommand{\dstzlnu}{D^{*0}\ell\bar{\nu}}
\newcommand{\bdstzlnu}{B^-\to D^{*0}\ell\bar{\nu}} 
\newcommand{\dstplnu}{D^{*+}\ell\bar{\nu}}
\newcommand{\bdstplnu}{\bar{B^0}\to D^{*+}\ell\bar{\nu}} 
\newcommand{\dstz}{D^{*0}}
\newcommand{\dstp}{D^{*+}}
\newcommand{\dst}{D^{*}}
\newcommand{\ddstlnu}{D^{**}\ell\bar{\nu}}
\newcommand{\bddstlnu}{\bar{B}\to D^{**}\ell\bar{\nu}} 
\newcommand{\dstxlnu}{D^* X\ell\bar{\nu}}
\newcommand{\bdstxlnu}{\bar{B}\to D^* X\ell\bar{\nu}} 
\newcommand{\dstpxlnu}{D^{*+} X\ell\bar{\nu}}
\newcommand{\dstzxlnu}{D^{*0} X\ell\bar{\nu}}
\newcommand{\vcb}{|V_{cb}|}
\newcommand{\vcbf}{|V_{cb}|{\cal F}(1)}
\newcommand{\fw}{{\cal F}(w)}
\newcommand{\fone}{{\cal F}(1)}
\newcommand{\bbbar}{B\bar{B}}
\newcommand{\Ufs}{\Upsilon(4S)}
\newcommand{\deltam}{\Delta m}
\newcommand{\mkpi}{m(K\pi)}
\newcommand{\mgg}{m(\gamma\gamma)}
\newcommand{\cby}{\cos\theta_{B-D^*\ell}}

\newcommand{\gevc}{GeV$/c$}
\newcommand{\gevcc}{GeV$/c^2$}
\newcommand{\mev}{MeV}
\newcommand{\mevc}{MeV$/c$}
\newcommand{\mevcc}{MeV$/c^2$}
\newcommand{\invps}{${\text{ps}^{-1}}$}
\newcommand{\invfb}{${\text{fb}^{-1}}$}
\newcommand{\etal}{\textit{et al.}}
\newcommand{\PRL}[3]{{Phys. Rev. Lett.} \textbf{#1}, #2 (#3)}
\newcommand{\PRD}[3]{{Phys. Rev. D} \textbf{#1}, #2 (#3)}
\newcommand{\PLB}[3]{{Phys. Lett. B} \textbf{#1}, #2 (#3)}
\newcommand{\NPB}[3]{{Nuc. Phys.} \textbf{B#1}, #2 (#3)}
\newcommand{\ZPC}[3]{{Z. Phys. C} \textbf{#1}, #2 (#3)}

\begin{document}

\preprint{CLNS 01/1774}
\preprint{CLEO 01-27}

\title{Determination of the $\bar{B} \rightarrow D^*\ell\bar{\nu}$ 
Decay Width and $|V_{cb}|$}  

\author{N.~E.~Adam}
\author{J.~P.~Alexander}
\author{C.~Bebek}
\author{B.~E.~Berger}
\author{K.~Berkelman}
\author{F.~Blanc}
\author{V.~Boisvert}
\author{D.~G.~Cassel}
\author{P.~S.~Drell}
\author{J.~E.~Duboscq}
\author{K.~M.~Ecklund}
\author{R.~Ehrlich}
\author{L.~Gibbons}
\author{B.~Gittelman}
\author{S.~W.~Gray}
\author{D.~L.~Hartill}
\author{B.~K.~Heltsley}
\author{L.~Hsu}
\author{C.~D.~Jones}
\author{J.~Kandaswamy}
\author{D.~L.~Kreinick}
\author{A.~Magerkurth}
\author{H.~Mahlke-Kr\"uger}
\author{T.~O.~Meyer}
\author{N.~B.~Mistry}
\author{E.~Nordberg}
\author{M.~Palmer}
\author{J.~R.~Patterson}
\author{D.~Peterson}
\author{J.~Pivarski}
\author{D.~Riley}
\author{A.~J.~Sadoff}
\author{H.~Schwarthoff}
\author{M.~R.~Shepherd}
\author{J.~G.~Thayer}
\author{D.~Urner}
\author{B.~Valant-Spaight}
\author{G.~Viehhauser}
\author{A.~Warburton}
\author{M.~Weinberger}
\affiliation{Cornell University, Ithaca, New York 14853}
% ...
%\author{(CLEO Collaboration)} %FOR PRD_SPECIAL_CHANGEME
\collaboration{CLEO Collaboration} %FOR PRL,CLNS
\noaffiliation

%\date{\today}
\date{October 14, 2002}

\begin{abstract}
In the Standard Model, the charged current of the weak interaction is
governed by a unitary quark mixing matrix that also leads to $CP$ violation.
Measurement of the Cabibbo-Kobayashi-Maskawa (CKM) matrix elements
is essential to searches for new physics, either through the
structure of the CKM matrix, or a departure from unitarity.
We determine the CKM matrix element $\vcb$ using a sample of
$3\times 10^6$ $B\bar{B}$ events in the CLEO detector at the Cornell
Electron Storage Ring. 
We determine the yield of 
reconstructed $\bdstplnu$ and $\bdstzlnu$ decays as a
function of $w$, the boost of the $D^*$ in the $B$ rest frame,
and from this we obtain
the differential decay rate $d\Gamma/dw$.  By extrapolating 
$d\Gamma/dw$ to $w=1$, the kinematic end point at which the $D^*$ is at rest 
relative to the $B$, we extract the product $\vcbf$, where $\fone$\ is
the form factor at $w=1$.  
We find $\vcbf = 0.0431\pm0.0013{\text{(stat.)}}\pm
0.0018{\text{(syst.)}}$. 
We combine $\vcbf$ with theoretical results for $\fone$ to determine 
$\vcb = 0.0469 \pm 0.0014{\text{(stat.)}} \pm 0.0020{\text{(syst.)}} \pm
0.0018{\text{(theo.)}}$.
We also integrate the differential decay rate over $w$ to
obtain ${\cal B}(\bdstplnu) = (6.09 \pm 0.19 \pm 0.40)\%$ and
${\cal B}(\bdstzlnu) = (6.50 \pm 0.20 \pm 0.43)\%$.

\end{abstract}

\maketitle

\section{Introduction}

The elements of the Cabibbo-Kobayashi-Maskawa (CKM) quark mixing matrix
\cite{Ckm,cKM} are fundamental parameters of the Standard Model and must
be determined experimentally. Measurement of the matrix elements tests
unified theories that predict the values of these elements.  It also
offers a means of searching for physics beyond the Standard Model by
testing for apparent deviations of the matrix from unitarity,
deviations that could arise if new physics 
affected the measurement of one of its elements.  The status of this test
is often displayed using the famous 
``Unitarity Triangle''\cite{unitaritytriangle}. The CKM matrix element
$\vcb$ sets the length of the base of this triangle, and it scales the 
constraint imposed by $\epsilon_K$ (This constraint scales as
$\vcb^4$), the parameter that quantifies $CP$ violation in the mixing
of neutral kaons \cite{K0cpv}. 

Two strategies are available for precise measurement of $\vcb$, both
of which rely on the underlying quark decay $b\to c\ell\bar{\nu}$,
where $\ell$ indicates $e^-$ or $\mu^-$.  
The first method combines measurements of the inclusive semileptonic
branching fraction and lifetime to determine the semileptonic decay
rate of the $B$ meson, which is proportional to $\vcb^2$.
Theoretical quark-level calculations give the proportionality
constant, thereby determining $\vcb$, with some uncertainties from
hadronic effects.
This first approach relies on the validity of quark-hadron duality,
the assumption 
that this inclusive sum is insensitive to the details of the various final
states that contribute. 

The second approach uses the specific decay mode $\bdstlnu$ or
$\bar{B}\to D\ell\bar{\nu}$. The rate for 
these decays depends not only on $\vcb$ and well-known weak decay physics, but 
also on strong interaction effects, which are parameterized by form factors.  
In general, these effects are notoriously difficult to quantify, but
because the $b$ and $c$ quark are both massive compared to the scale
of hadronic physics, $\bar\Lambda\approx 0.5$ GeV, heavy-quark symmetry
relations can be applied to $\bar{B}\to D^{(*)}\ell\bar{\nu}$ decays
\cite{hqs_voloshin,hqs_isgur,hqs_luke,hqs_falk,hqs_neubert}.
In the limit $m_b,m_c\to\infty$, the form factor is unity at zero
recoil, the kinematic point at which the final state $D^{(*)}$ is a rest
with respect to the initial $B$ meson.  Corrections to the
infinite-mass limit are then calculated using an expansion in powers of
$\bar\Lambda/m_Q$.  Luke showed \cite{hqs_luke} that the first-order
correction vanishes for pseudoscalar-to-vector transitions, making
$\dstlnu$ decays more attractive theoretically than $D\ell\bar{\nu}$
for $\vcb$ determination.\footnote{
There are experimental advantages as well: a larger branching fraction,
a distinctive final state with the narrow $D^*$, and 
less phase-space suppression than the $P$-wave decay 
$\bar B\to D\ell\bar{\nu}$ near the important zero-recoil point.}
Heavy Quark Effective Theory
(HQET)~\cite{hqet_grinstein,hqet_eichten,hqet_georgi,hqet_koerner,hqet_mannel}
exploits the heavy-quark symmetry and offers a rigorous framework for
quantifying the hadronic effects with relatively small
uncertainty~\cite{dslnu_neubert,dslnu_falk}.

In this paper, we report more fully on a recently published~\cite{letter}
measurement of $\vcb$ using $\bdstlnu$ decays
that are detected in the CLEO II detector at the Cornell Electron
Storage Ring (CESR).  The $\bdstlnu$ decays
are fully reconstructed, apart from the neutrino.  The analysis takes 
advantage of the kinematic constraints available at the $\Ufs$
resonance, where the data were collected, to suppress backgrounds, help 
distinguish $\dstlnu$ from similar modes such as $D_1\ell\bar{\nu}$,
and provide  
superb resolution on the decay kinematics. This analysis is the first since
a previous CLEO result~\cite{oldcleo} to use not only
$\bar{B}^0 \to D^{*+}\ell \bar{\nu}$ decays, but also
$B^- \to D^{*0}\ell \bar{\nu}$ decays~\cite{blvthesis}.  Consistency
between these two modes is a valuable cross-check of our results.

We reconstruct $\dstp$ candidates and their charge conjugates
(charge conjugates are implied throughout this paper) through the 
modes $\dstp\to D^0 \pi^+$ and $D^0\to K^-\pi^+$, and we reconstruct 
$\dstz$ candidates through the modes $\dstz\to D^0 \pi^0$, $D^0\to K^-\pi^+$, 
and $\pi^0\to \gamma\gamma$.  Each $D^*$ candidate is combined with an
electron or muon candidate. We then divide the reconstructed candidates into bins of 
$w$, where $w$ is the scalar product of the $B$ and $D^*$ four-velocities,
and equals the relativistic $\gamma$ of the $D^*$ in the $B$ rest 
frame.\footnote{The variable $w$ is linearly related to $q^2$, 
the squared invariant
mass of the virtual $W$, via $w=(m_B^2 + m_{D^*}^2 - q^2)/(2m_B m_{D^*})$, 
where $m_B$ and $m_{D^*}$ are the $B$- and $D^*$-meson masses.}
Given these yields as a
function of $w$, we fit simultaneously for parameters describing 
the form factor $\fw$ and the normalization at $w=1$. This normalization is 
proportional to the product $\vcb^2{\cal F}^2(1)$, and
combined with the theoretical results for $\fone$, it gives us $\vcb$.

\section{Event Samples}
Our analysis uses $3.33 \times 10^6$ $\bbbar$ events (3.1 \invfb) produced 
on the $\Ufs$ resonance at the Cornell Electron Storage Ring (CESR) and 
detected in the CLEO II detector. In addition, the analysis 
uses a sample of 1.6 \invfb\ of data collected slightly below the 
$\Ufs$ resonance for the purpose of subtracting continuum backgrounds.
Because of miscalibration of low-energy showers in the calorimeter in a 
subset of the data, we
use only $3.04 \times 10^6$ $\bbbar$ events (2.9 \invfb) produced on the
$\Ufs$ resonance and 1.5 \invfb\ of data collected below the 
$\Ufs$ resonance for reconstructing $\bdstzlnu$ candidates.

The CLEO II detector~\cite{cleonim} has three central tracking
chambers, immersed in a 1.5 T magnetic field, that measure charged
particle trajectories and momenta.  The momentum resolution is 5
\mevc\ (12 \mevc) for particles with a momentum of 1 \gevc\ (2
\gevc) (typical for the lepton and the $K$ and $\pi$ from the $D^0$)
and 3 \mevc\ for particles with momentum less than 250 \mevc\ 
(typical for the $\pi^+$ from the $\dstp$).  A CsI(Tl) calorimeter
surrounds both the tracking chambers and a time-of-flight system that
is not used for this analysis.  The calorimeter provides photon
detection and assists with electron identification. The energy
resolution of the calorimeter is 3.8 MeV for 100 MeV photons, a
typical energy for photons from the decay of the $\pi^0$ from the $\dstz$
decay. The outermost detector component consists of plastic streamer
counters layered between iron plates and provides detection of muons.

We also use simulated event samples from a \textsc{Geant}-based~\cite{geant}
Monte Carlo simulation.  With this Monte Carlo, we produce large
samples of simulated $\bdstlnu$ decays as well as a sample of
$16\times 10^6$ $\bbbar$ events to study some backgrounds.

\section{Event Reconstruction}
The $\Ufs$ is produced and decays at rest, and each daughter $B$ meson
is produced with a momentum of about 0.3 GeV$/c$.  As a result, 
$\Ufs\to\bbbar$ events tend to be isotropic, or ``spherical,'' 
with particles carrying energy in all directions.  
When the electron-positron collisions in CESR do
not produce $\Ufs$'s, they can produce, among other things, 
$q{\bar q}$ quark pairs, where the $q$ is a $c$, $s$, $u$, or $d$ quark. 
Because the mass of these quark pairs is much lower than the energy of the 
beam, the daughter particles of these quarks' hadronization have higher 
momenta than the $B$'s. These events tend to have a more ``jetty'' appearance; 
that is, the energy in the event tends to be distributed back-to-back. The 
ratio of Fox-Wolfram moments $H_2/H_0$~\cite{r2} measures an event's
jettiness, with values of the ratio approaching zero for spherical
events, and approaching one for jetty events.
To suppress non-$\bbbar$ events, we 
require that the ratio of Fox-Wolfram moments $H_2/H_0$ be less than 
0.4, a condition satisfied by 98\% of $B{\bar B}$ events containing a 
$\dstlnu$ decay.

\begin{figure*}
\epsfig{file=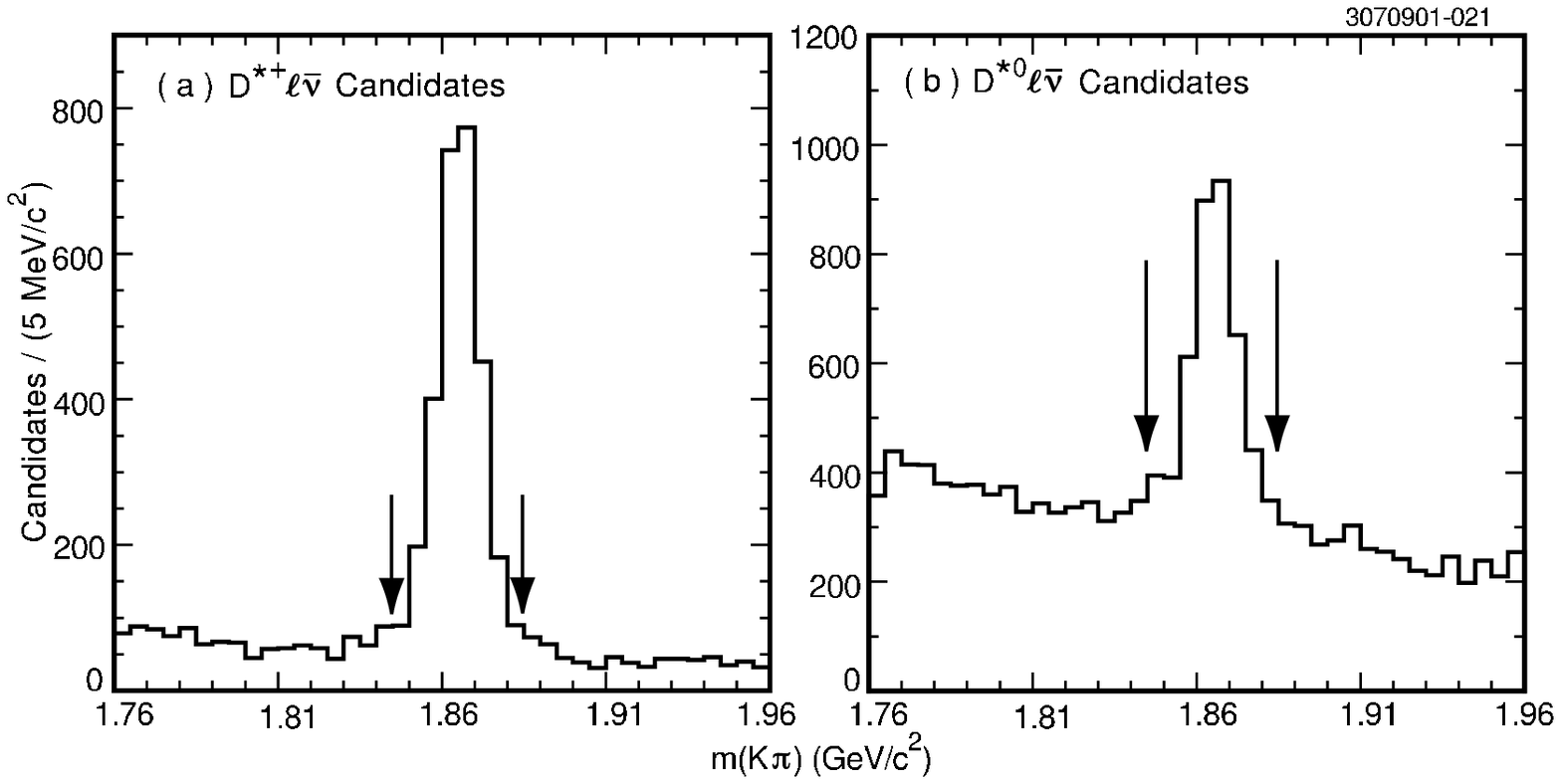,width=165mm}
\caption{\label{fig:mkpi}
The $\mkpi$ distribution for (a) $\dstplnu$ and (b) $\dstzlnu$ 
candidates. All requirements are met except $|m(K\pi)-1.865| \le 
0.020$ \gevcc. We accept candidates that fall between the arrows.}
\end{figure*}

To reconstruct $\dstlnu$ candidates we first form $D^0\to K^-\pi^+$
candidates from all possible pairs of oppositely charged tracks,
alternately assigning one the kaon mass and the other the pion mass. 
We require a fiducial cut of $|\cos\theta|\le 0.9$ for tracks, where
$\theta$ is the polar angle of the track's momentum vector with respect 
to the $e^+ e^-$ beam axis. Tracks outside this fiducial region 
are excluded from consideration because they are poorly measured,
having passed through the 
endplate of one of the inner tracking chambers and therefore either
traversing a significant amount of material before entering the outer
tracking chamber or never entering it at all. 
We reconstruct the invariant mass
$\mkpi$ of the $D$ 
candidate with a resolution of about 7 \mevcc, accepting candidates that
lie in the window $|m(K\pi)-1.865| \le 0.020$ \gevcc.  The $\mkpi$
distributions for $\dstplnu$ and $\dstzlnu$ candidates are shown in
Fig.~\ref{fig:mkpi}. 

\begin{figure*}
\epsfig{file=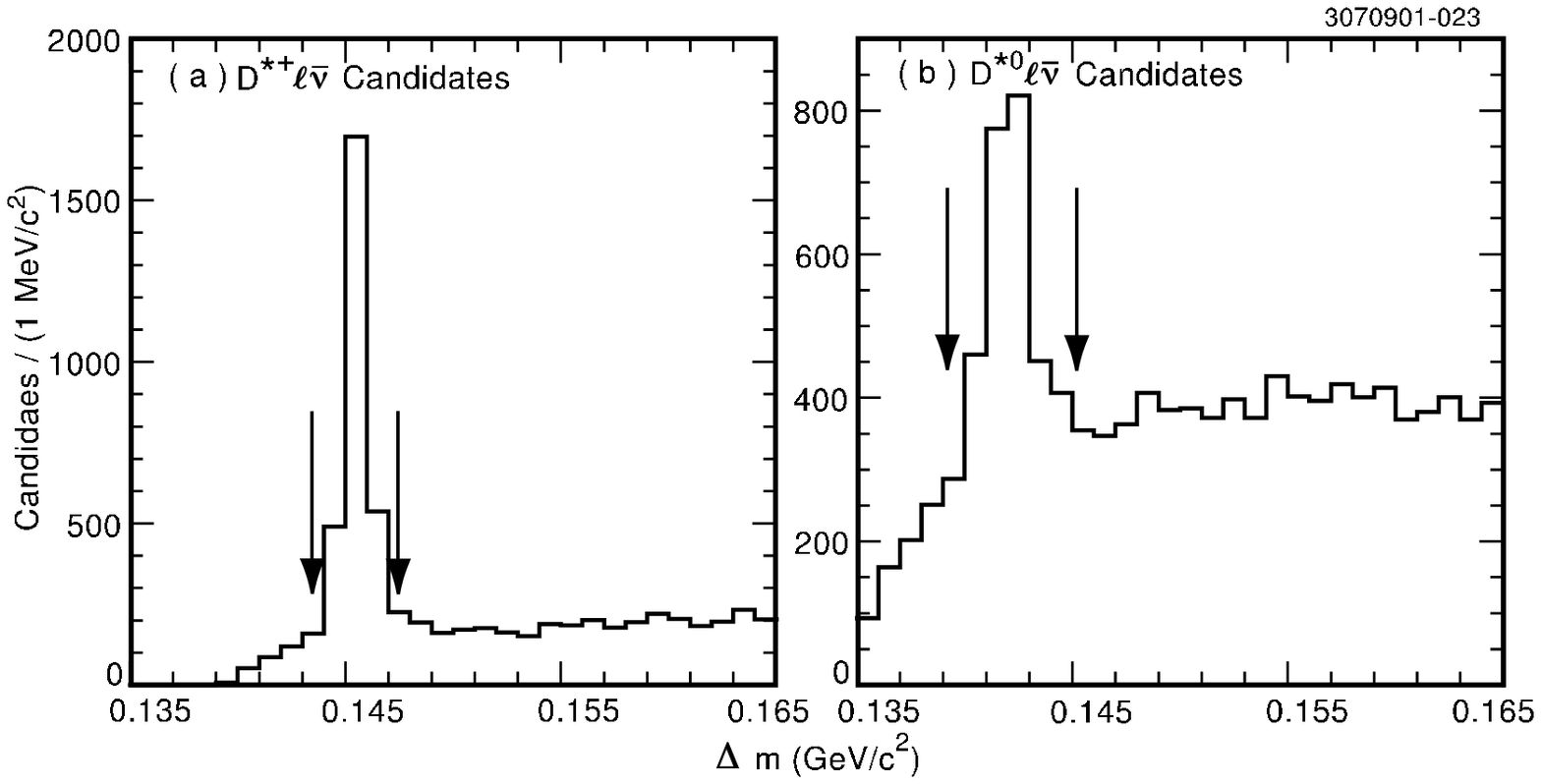,width=165mm}
\caption{\label{fig:dm_dstp}
The $\deltam$ distribution for (a) $\dstplnu$ and (b) $\dstzlnu$
candidates. All requirements are met except $|\deltam- 0.14544|\le
0.002$ \gevcc\ for $\dstplnu$ and $|\deltam - 0.1422|\le 0.003$ \gevcc\ for
$\dstzlnu$. We accept candidates that fall between the arrows.}
\end{figure*}

The pions produced in the decay $\dst\to D \pi$ have low momentum
($<250$ \mevc) 
because the combined mass of the $D^0$ and $\pi$ is within 8 \mevcc\ of
the mass of the $\dst$. We label these pions ``slow.''
For $\dstp$ candidates, we add a slow $\pi^+$ candidate to a $D^0$ candidate,
requiring that the slow pion have the same charge as the pion from the
$D$ decay.
This pion must also satisfy $|\cos\theta| \le 0.9$. 
The $K$ and $\pi$ are fit to a common vertex, and then the slow $\pi^+$ and 
$D^0$ are fit to a second vertex using the beam spot constraint. 
For this vertexing we use error matrices from our Kalman fitter~\cite{Kalman}.
We then form $\deltam\equiv m(K\pi\pi)- m(K\pi)$.  
We look at $\deltam$ rather than $m(K\pi\pi)$
because subtracting the $D^0$ candidate mass from the $D^*$ candidate mass
cancels some errors in reconstructing the $D^0$. 
A plot of $\deltam$ for $\dstp$ 
candidates is shown in Fig.~\ref{fig:dm_dstp}(a). The 
vertex constraints improve the resolution by about 20\% to 0.7 \mevcc.
We require  $|\deltam - 0.14544|\le 0.002$ \gevcc\ for $\dstp$ candidates.

\begin{figure}
\epsfig{file=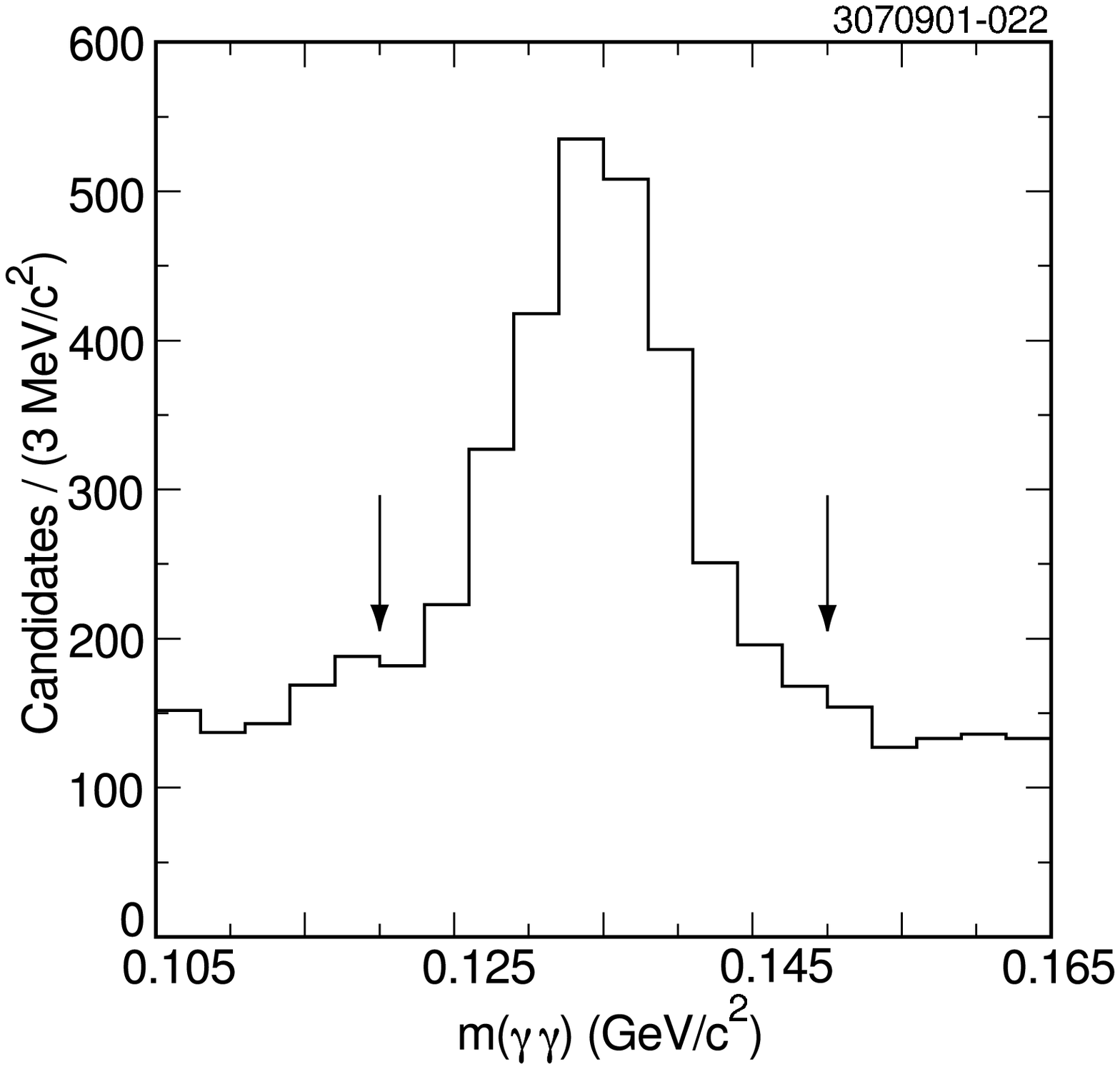,width=3.4in}
\caption{\label{fig:mgg_dm_dstz}
The $\mgg$ distribution for $\dstzlnu$ candidates.
All requirements are met except $0.120$ \gevcc\ $\le \mgg \le 0.150$
\gevcc. We accept candidates that fall between the arrows.}
\end{figure}

For $\dstz$ candidates, we add a slow $\pi^0\to\gamma\gamma$ candidate
to the $D^0$ candidate. We  
construct $m(\gamma\gamma)$ for slow $\pi^0$ candidates from showers in the 
CsI calorimeter whose position is inconsistent with extrapolation of any of 
the tracks reconstructed in the event. We require that the lateral pattern
of energy deposition in the calorimeter
be consistent with expectations for a photon. Particles 
with $|\cos\theta|>0.71$ travel through the endplate of the outermost 
tracking chamber before reaching the calorimeter, again traversing a 
significant amount of material. We therefore require that both photon
candidates satisfy $|\cos\theta|\le 0.71$ so as to remain in the part
of the calorimeter  with the best energy and position resolution. Both
photons must have energy greater than 30 \mev\ to limit background from
soft showers. We also  require the invariant mass $\mgg$ to give
the known $\pi^0$ mass within roughly three times the resolution of 5
\mevcc: $0.120$ \gevcc\ $\le \mgg \le 0.150$ \gevcc.
The $\deltam$ resolution for $\dstz$'s is about 0.9 \mevcc, so we require
$|\deltam - 0.1422|\le 0.003$ \gevcc. The $\deltam$ distribution for
$\dstz$ candidates is shown in Fig.~\ref{fig:dm_dstp}(b), and the $\mgg$ 
distribution is shown in Fig.~\ref{fig:mgg_dm_dstz}.

Finally, we require the momentum of the $D^*$ candidate to be less than 
$\frac{1}{2}\sqrt{E_B^2-m(K\pi\pi)^2}$, (approximately 2.5~\gevc), 
where $E_B$ is the energy of the beam. This requirement suppresses
background from  non-$\bbbar$ events.

We next combine the $D^*$ candidate with a lepton candidate, accepting both 
electrons and muons.
Electrons are identified using the ratio of their energy deposition in
the CsI calorimeter to the reconstructed track momentum, the shape
of the shower in the calorimeter, and their specific ionization in the
tracking chamber.  We require our candidates to lie in the momentum range 
$0.8$~\gevc\ $\le p_e < 2.4$~\gevc, where
the upper bound is the end point of $\dstlnu$ decays.  
This momentum selection is approximately 93\% efficient for ${\bar
B}\to D^* e^-{\bar \nu}$ decays.
We require muon candidates to penetrate two layers of steel in the solenoid 
return yoke, or about 5 interaction lengths. Only muons with momenta 
above about 1.4 \gevc\ satisfy this requirement; we therefore demand
that muon candidates lie in  
the momentum range $1.4$~\gevc\ $\le p_e < 2.4$ \gevc.  
This more restrictive muon momentum requirement has an efficiency of
approximately 61\%. We require both
muon and electron candidates to be in the central region of the
detector ($|\cos\theta|\le 0.71$), where efficiencies and hadron
misidentification rates are well understood. The charge of the lepton
must match the charge of the  kaon, and in the case of $\dstplnu$
decays, be opposite that of the slow pion.

The remaining reconstruction relies on the kinematics of the 
$\bdstlnu$ decay.  We first reconstruct $\cby$, the
angle between the $D^*$-lepton 
combination and the $B$ meson, computed assuming that the only 
unreconstructed particle is a neutrino.  This variable helps
distinguish $\bdstlnu$ decays from background and is
necessary for the reconstruction of $w$.
 To form $\cby$, we first
note that the 4-momenta of the particles involved in $\bdstlnu$ decay
are related by
\begin{equation}
p_{\nu}^{2} = (p_{B} - p_{D^*} - p_\ell)^{2}.
\end{equation}
Setting the neutrino mass to zero gives
\begin{equation}
0=m_{B}^{2} + m(D^{*}\ell)^{2} - 2\left[E(B)E(D^{*}\ell) - {\mathbf p}(B)
\cdot {\mathbf p}(D^{*}\ell)\right].
\end{equation}
We solve for the only unknown quantity, the
angle between the $B$ meson and the $D^*$-lepton pair:
\begin{equation}
\cby = \frac{2E(B) E(D^*\ell) - m_B^2 - m(D^*\ell)^2}{2|{\mathbf p}(B)||{\mathbf p}(D^*\ell)|}.
\label{eq:cby}
\end{equation}
In forming $\cby$, we use the momenta of the $\dst$ and lepton
candidates as well as the $B$ mass~\cite{ershov} and average $B$
momentum, measured in our data.  At CESR, a symmetric $e^+e^-$
collider operating on the $\Ufs$ resonance, the $B$ energy and
therefore momentum is given by the energy of the colliding beams.
Instead of relying on beam energy measurements based on storage ring
parameters and subject to significant uncertainties, we determine the
average $B$ momentum directly using fully reconstructed $B$ decays to
hadrons.  The energy spread of the beams and run-to-run energy
variations lead to a distribution of $B$ energies and momenta.  By
measuring the momentum distribution of fully reconstructed hadronic
$B$ decays in our data sample, we determine the energy spread
intrinsic to CESR, which is then used to simulate $\bbbar$ pair
production in our Monte Carlo.  For $\cby$ we use the true $\dst$ mass
rather than the reconstructed $m(K\pi\pi)$ to avoid a bias in the
$\cby$ distribution of the $\deltam$ sideband, which we use to
determine a background.

\begin{figure*}
\epsfig{file=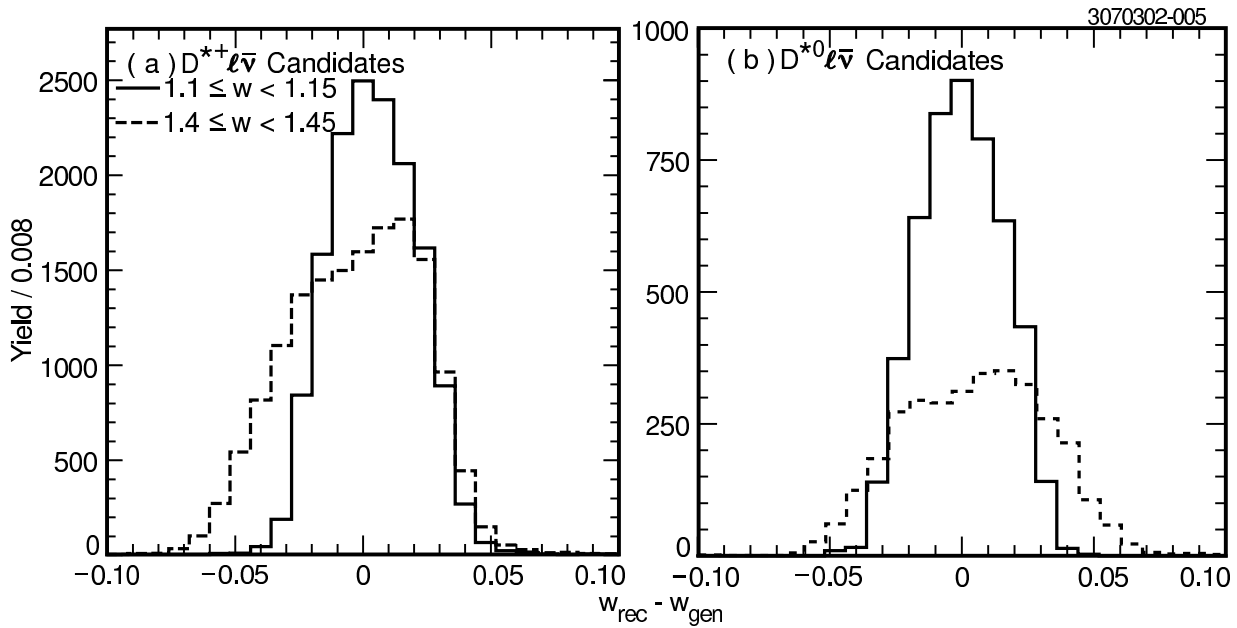,width=165mm}
\caption{\label{fig:wrec}
The difference between the reconstructed $w$ and generated
$w$ for simulated $\dstplnu$ (a) and $\dstzlnu$ (b) decays in the
intervals $1.1 \le w < 1.15$ (solid) and $1.4 \le w < 1.45$ (dashed).
The normalization of all four histograms is arbitrary.}
\end{figure*}

We next estimate $w$ for each candidate.  Exact reconstruction of
$w$, the boost of the $D^*$ in the rest frame of the $B$,
requires knowledge of the $B$ momentum vector.
Although the magnitude of the $B$ momentum is known, the $B$
direction is unknown.
However, it must lie on a cone with opening angle
$\theta_{B-D^*\ell}$ around the $D^*\ell$ direction.
We calculate $w$ for all $B$ flight directions on this cone and
average the smallest and largest values to estimate $w$, with typical
resolution of 0.03.  
We divide our sample into ten equal bins from 1.0 to 1.5, where the
upper bound is just below the kinematic limit of 1.504.  For a few candidates, 
the reconstructed $w$ falls outside our range; we 
assign these to the first or last bin as appropriate. 
Figure~\ref{fig:wrec} shows the distributions of 
reconstructed $w$ minus generated $w$ in the third and ninth $w$ bins for
simulated $\bdstplnu$ and $\bdstzlnu$ decays.

\begin{figure}
\epsfig{file=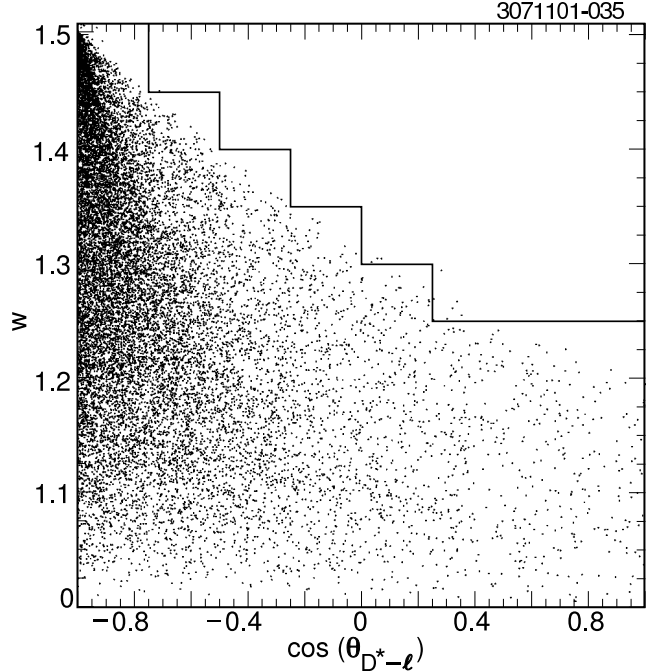,width=3.4in}
\caption{\label{fig:cdslep_v_w}
The distribution of $w$ versus $\cos\theta_{D^*-\ell}$ for simulated
$\dstlnu$ decays with lepton momentum between 
0.8 \gevc\ $\le p_{\ell} <$~2.4 \gevc. We accept candidates that fall below 
and to the left of the stair-step line.}
\end{figure}

In the high $w$ bins, 
we suppress background with minor loss of signal efficiency by 
restricting the cosine of the angle between the momenta of the $D^*$
and of the lepton 
($\cos\theta_{D^*-\ell}$). The distribution of $\cos\theta_{D^*-\ell}$ versus
$w$ is shown in Fig.~\ref{fig:cdslep_v_w} for simulated $\bdstlnu$ 
decays.  Some backgrounds are uniformly distributed in this angle.  The
accepted angles are listed in Table~\ref{tab:cdslep}.

\begin{table}
\caption{\label{tab:cdslep}
The accepted regions of the cosine of the angle between the $D^*$
and the lepton in each $w$ bin.}
\begin{ruledtabular}
\begin{tabular}{lcdd}
$w$ bin & $w$ limits & \multicolumn{2}{c}{Accepted $\cos\theta_{D^*-\ell}$} \\
&  & \multicolumn{1}{c}{min.}  & \multicolumn{1}{c}{max.} \\ \hline
1-5   & $<1.25$   & -1.00 &  1.00 \\
6     & 1.25-1.30 & -1.00 &  0.25 \\
7     & 1.30-1.35 & -1.00 &  0.00 \\
8     & 1.35-1.40 & -1.00 & -0.25 \\
9     & 1.40-1.45 & -1.00 & -0.50 \\
10    & $\ge 1.45$& -1.00 & -0.75 \\
\end{tabular}
\end{ruledtabular}
\end{table}

\section{Extracting the $\dstlnu$ Yields}
\subsection{Method}
\begin{figure}
\epsfig{file=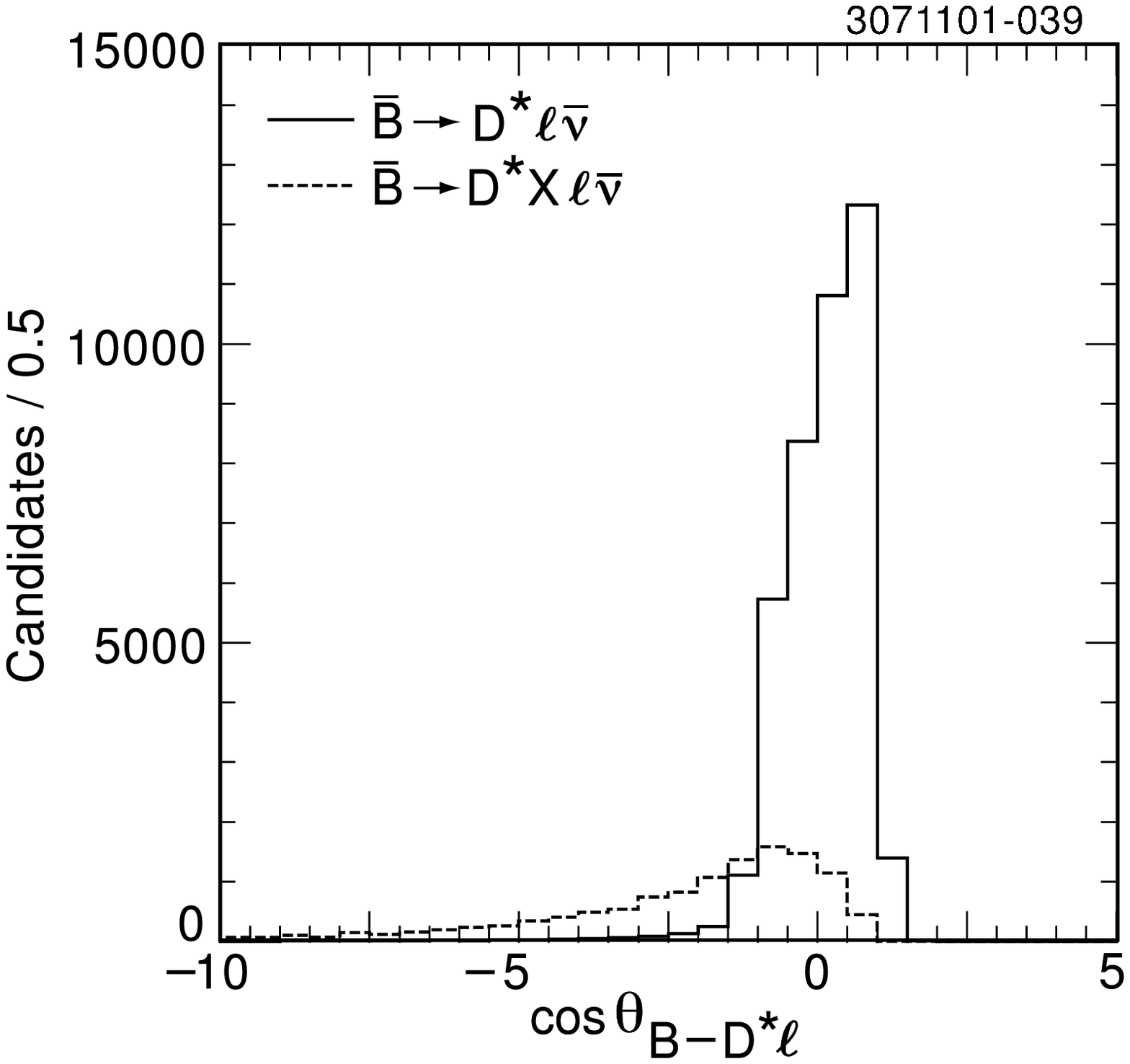,width=3.4in}
\caption{\label{fig:cbysig}
The $\cby$ distributions for simulated $\dstlnu$ and $\dstxlnu$ decays.} 
\end{figure}

At this stage, our sample of candidates contains not only $\bdstlnu$
decays, but also $\bddstlnu$ and $\bar{B}\to D^*\pi\ell\bar{\nu}$ 
decays and various backgrounds.  In the following, we refer 
to $\bddstlnu$ and non-resonant $\bar{B}\to D^*\pi\ell\bar{\nu}$
decays collectively as $\bdstxlnu$ decays.  
In order to disentangle the $\dstlnu$ from the $\dstxlnu$ decays, we use a 
binned maximum likelihood fit~\cite{likelihood} to the $\cby$ distribution. 
As shown in Fig.~\ref{fig:cbysig}, $\bdstlnu$ decays are concentrated 
in the physical region, $-1 \le \cby \le 1$, while the missing mass of 
the $\dstxlnu$ decays allows them to populate $\cby < -1$. In this fit, 
the normalizations of the various background distributions are fixed and we 
allow the normalizations of the $\dstlnu$ and the $\dstxlnu$ components
to float. For each $w$ bin, we fit over a $\cby$ region chosen to
include 95\% of the  $\dstxlnu$ events in that bin. These regions are
listed in Table~\ref{tab:fitregion}.

\begin{table}
\caption{\label{tab:fitregion}
The regions of $\cby$ over which we perform a binned maximum 
likelihood fit.}
\begin{ruledtabular}
\begin{tabular}{lcdd}
$w$ bin  & $w$ limits & \multicolumn{2}{c}{$\cby$ fit region} \\
&           & \multicolumn{1}{c}{min.} & \multicolumn{1}{c}{max.}\\ \hline
1-6   & $<1.30$   & -8.0 & 1.5 \\
7     & 1.30-1.35 & -6.0 & 1.5 \\
8     & 1.35-1.40 & -4.0 & 1.5 \\
9     & 1.40-1.45 & -3.0 & 1.5 \\
10    & $\ge 1.45$& -2.0 & 1.5 \\
\end{tabular}
\end{ruledtabular}

\end{table}

The distributions of the $\dstlnu$ and $\dstxlnu$ decays come from Monte
Carlo simulation.  We simulate $\dstlnu$ decays using the form factor
of~\cite{caprini} and include the effect of final-state radiation
($\bdstlnu\gamma$) using \textsc{Photos}~\cite{photos}.
For $\dstxlnu$, we model $\ddstlnu$ modes according
to ISGW2~\cite{isgw2} and non-resonant $D^*\pi\ell\bar{\nu}$ from Goity and
Roberts~\cite{goityroberts}.  Our model for $\dstxlnu$ is dominated by
approximately equal parts of $D_1 \ell\bar{\nu}$ and
$D^*\pi\ell\bar{\nu}$.  The other backgrounds, and how we obtain
their $\cby$ distributions and normalizations, are described in the
next section. 

\subsection{Backgrounds}
There are several sources of decays other than $\bdstlnu$ and 
$\bdstxlnu$. We divide these backgrounds into five classes: continuum, 
combinatoric, uncorrelated, correlated, and fake lepton. 
As an indication of the relative importance of the various
backgrounds, in Table~\ref{tab:bkgd_frac} we give both the fraction
${B_i}/{N_{\text{tot}}}$ of candidates from each background source
relative to all 
candidates and the ratio $B_i/S$ of each background source to $\dstlnu$
signal.  Because signal events populate the physical region
$-1\le\cby\le 1$, we compute both ${B_i}/{N_{\text{tot}}}$ and
$B_i/S$ using only candidates in this ``signal region.''
We discuss each background and how we determine it below.

\begin{table}
\caption{\label{tab:bkgd_frac}
The contribution of each background $B_i$ to the
total number of candidates $N_{\text{tot}}$ in the range $-1 \le \cby \le 1$ 
for the $\dstplnu$ and $\dstzlnu$ analyses.
The relative size to $\dstlnu$ signal $B_i/S$ is also given for the
same $\cby$ interval.}
\begin{ruledtabular}
\begin{tabular}{ldddd}
Background & \multicolumn{2}{c}{$\dstplnu$ Contribution} & 
             \multicolumn{2}{c}{$\dstzlnu$ Contribution}\\
& \multicolumn{1}{c}{${B_i}/{N_{\text{tot}}}$ (\%)} 
& \multicolumn{1}{c}{$B_i/S$  (\%)}
& \multicolumn{1}{c}{${B_i}/{N_{\text{tot}}}$ (\%)}
& \multicolumn{1}{c}{$B_i/S$  (\%)}\\
\hline
Continuum    & 3.8 &  4.7  &  2.8 &  5.1 \\
Combinatoric & 7.9 & 10    & 38   & 70   \\
Uncorrelated & 4.4 &  5.6  &  4.7 &  8.6 \\
Correlated   & 0.4 &  0.5  &  0.1 &  0.2 \\
Fake Lepton  & 0.5 &  0.6  &  0.2 &  0.4 \\
\end{tabular}
\end{ruledtabular}
\end{table}

\subsubsection{Continuum background}
At the $\Ufs$ we detect not only resonance events 
($\Ufs\to \bbbar$), but also 
non-resonant events such as $e^+e^- \to q\bar{q}$. This background 
contributes about 4\% of the candidates within the signal region for $\dstplnu$ 
decays, and about 3\% for $\dstzlnu$ decays. 
This is about 5\% relative to the $\dstlnu$ signal.
In order to subtract background from this source, CESR runs one-third of the 
time slightly below the $\Ufs$ resonance. For this continuum 
background, we use the $\cby$ distribution of candidates in the off-resonance data 
scaled by the ratio of luminosities and corrected for the small difference in 
the $e^+e^-\to q{\bar q}$ cross sections at the two center-of-mass
energies. In reconstructing 
$\cby$, we scale the energy and momentum of the $D^*$ and 
lepton by the ratio of the center-of-mass energies and
use the $B$ momentum measured in on-resonance data to compute 
the $B$ energy. This continuum background includes combinatoric
and fake lepton backgrounds arising from continuum processes.

\subsubsection{Combinatoric background}
\label{sec:comb}
 
Combinatoric background candidates are those in which one or more of 
the particles in the $D^*$ candidate does not come from a true $D^*$ decay.
This background contributes 8\% of the candidates in the signal region for
$\dstplnu$; for $\dstzlnu$, which suffers from random shower 
combinations and does not benefit from the
charge correlation of the slow pion, this background contributes 38\%
of the candidates in the signal region.  Relative to the $\dstlnu$ signal,
the combinatoric background is 10\% for $\dstplnu$ and 70\% for
$\dstzlnu$.

The $\cby$ distribution of the combinatoric background is provided by
$D^*$-lepton combinations in the high $\deltam$ sideband.
We choose the $\deltam$ sidebands of $0.155$~\gevcc\ 
$\le \deltam < 0.165$~\gevcc\ for $\dstplnu$ 
and $0.147$~\gevcc\ $\le \deltam < 0.165$~\gevcc\ for $\dstzlnu$.  For 
values of $\deltam$ above these ranges, the slow pions tend to be faster, and
therefore $\cby$ tends to be larger, while regions closer to the
$\deltam$ signal region include signal decays in which the
slow pion is 
poorly reconstructed.
With this choice, only 3.5\% and 0.4\%
of the $\dstlnu$ decays fall in the sideband for $\dstplnu$ and $\dstzlnu$, 
respectively.

\begin{figure*}
\epsfig{file=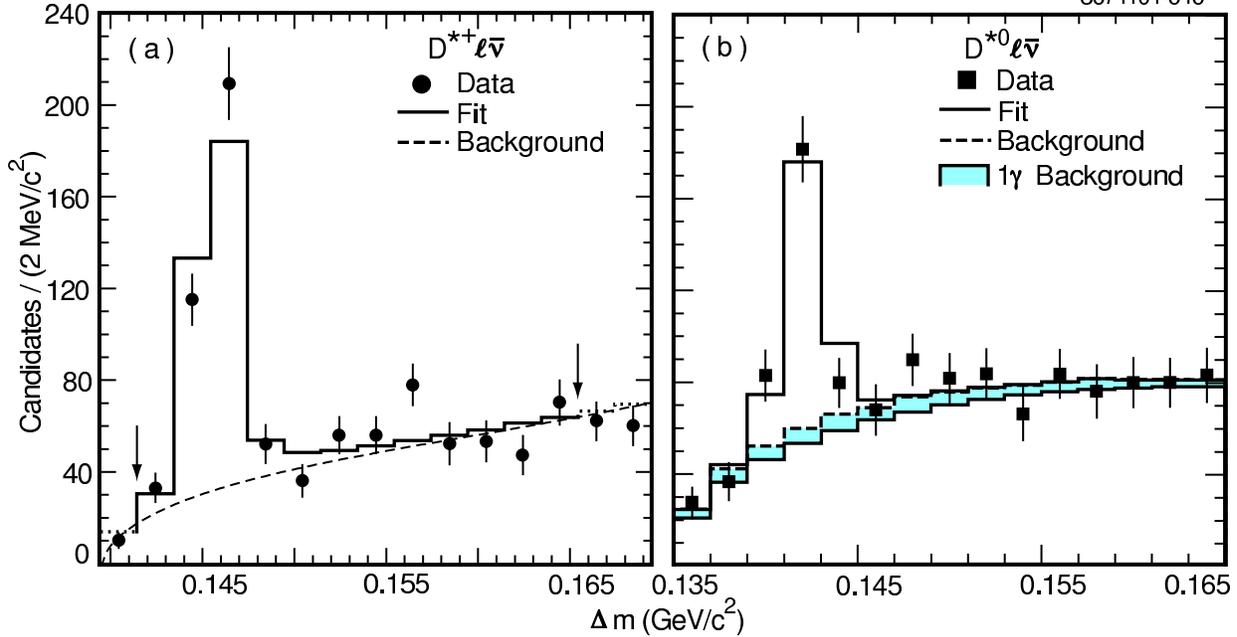,width=165mm}
\caption{ \label{fig:deltamfit}
The $\deltam$ distribution of candidates in the third $w$ bin
($1.1\le w < 1.15$) for (a) $\dstplnu$ candidates and (b) $\dstzlnu$
candidates with the result of the fit superimposed. The data (solid
circles or squares) are superimposed 
with the combinatoric background distribution 
(dashed curve) and the sum of the background and the $\dst$ signal 
(solid histogram).
In (a) the arrows delimit the fit region.  In (b), the
shaded histogram shows combinations in which only one of the two
photons forming the $\pi^0$ candidate was correct.
Unless indicated otherwise, the error bars provided in all figures are
statistical only.
}
\end{figure*}

\begin{figure*}
\epsfig{file=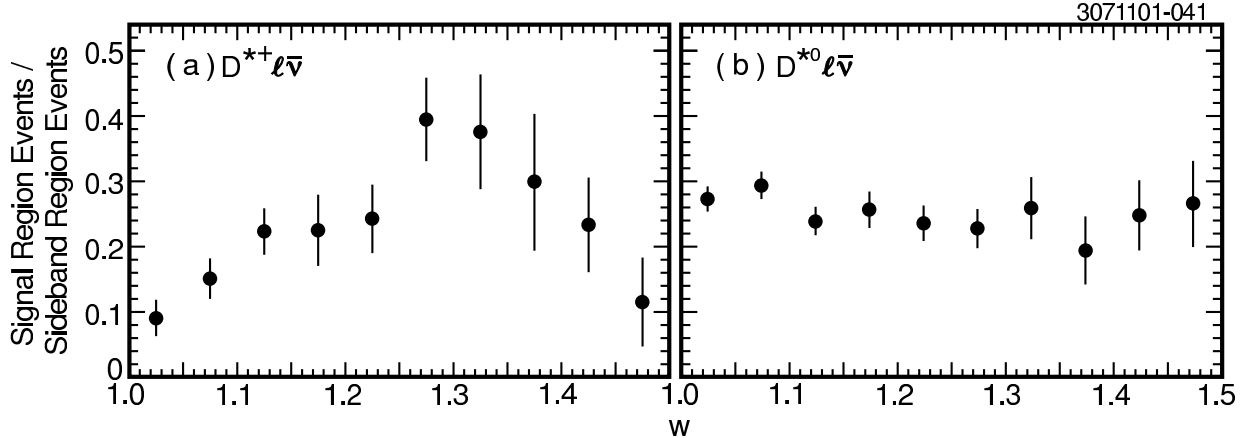,width=165mm}
\caption{\label{fig:combsf}
The ratio of the number of combinatoric background candidates in
the $\deltam$ signal region to the number in the $\deltam$ sideband, as
determined from fits to $\deltam$ for (a) $\dstplnu$ and (b) $\dstzlnu$.
The $\cby$ distribution of the combinatoric background is provided by
the sideband candidates normalized by this ratio.}
\end{figure*}

The normalization of the $\deltam$ sideband candidates is determined 
in each $w$ bin from a fit to the $\deltam$ distribution with
the sum of properly reconstructed $D^*$'s and the combinatoric background.
The line-shape for the $D^*$ peak is taken from simulated $\dstlnu$ decays. 
The $\dstzlnu$ line-shape includes $D^{*0}$ candidates 
in which only one of the two photons constituting the $\pi^0$ was correct.
Since these candidates preferentially populate the $\deltam$ signal region,
a few (3.9\% of all $\dstzlnu$ decays) remain after our 
combinatoric background subtraction and are
included in our $\dstzlnu$ signal. 
For $\dstplnu$ we assume 
a background shape of the form
\begin{equation}
n(\deltam - m_{\pi})^a \exp\left\{\left[c_1(\Delta m - m_{\pi})+
c_2(\Delta m - m_{\pi})^2\right]\right\} ,
\end{equation}
where $c_1$ and $c_2$ are constants fixed
using an inclusive $D^{*+}$ sample,
and we vary $n$, $a$, and the normalization of the signal peak. 
For $\dstzlnu$ we assume 
a background shape of the form 
\begin{equation}
n(\Delta m - m_{\pi})^{a} \exp\left[{b(\Delta m - m_{\pi})}\right]
\end{equation}
and vary $n$, $a$, $b$, and the normalization of the signal peak. 
The fits for $\dstplnu$ and $\dstzlnu$ are shown for a representative
$w$ bin in Fig.~\ref{fig:deltamfit}. The normalizations are shown in 
Fig.~\ref{fig:combsf}.

\begin{figure*}
\epsfig{file=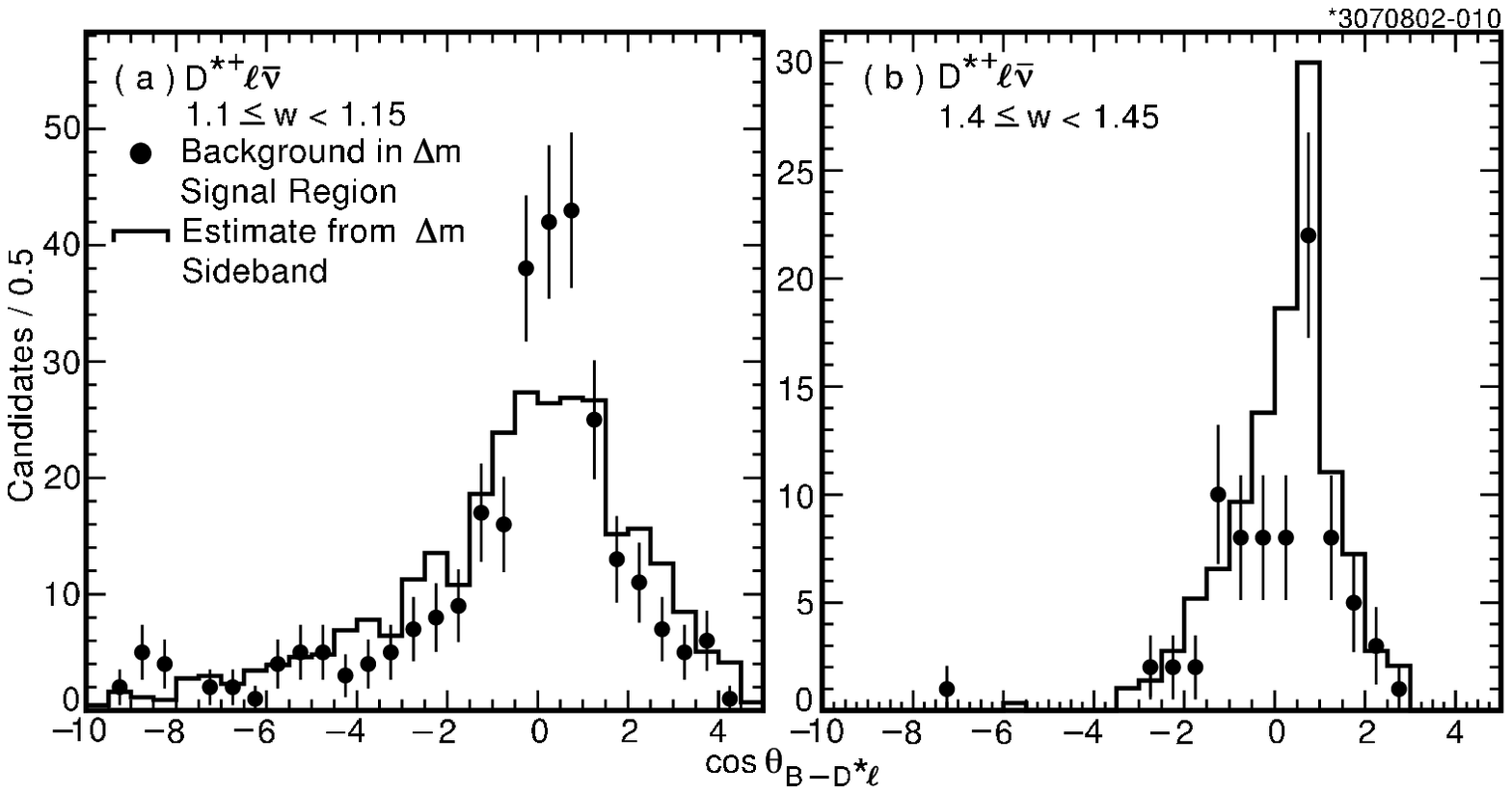,width=165mm}
\caption{\label{fig:comb_dsp}
From Monte Carlo, the $\cby$ distribution for $\dstplnu$ 
combinatoric background candidates in the $\deltam$ signal region (points)
and for scaled candidates from the $\deltam$ sideband (histogram) for 
(a) $1.1 \le w < 1.15$ and (b) $1.4 \le w < 1.45$.  The sideband events
are normalized using a fit to the $\deltam$ distribution as described
in the text.}
\end{figure*}

\begin{figure*}
\epsfig{file=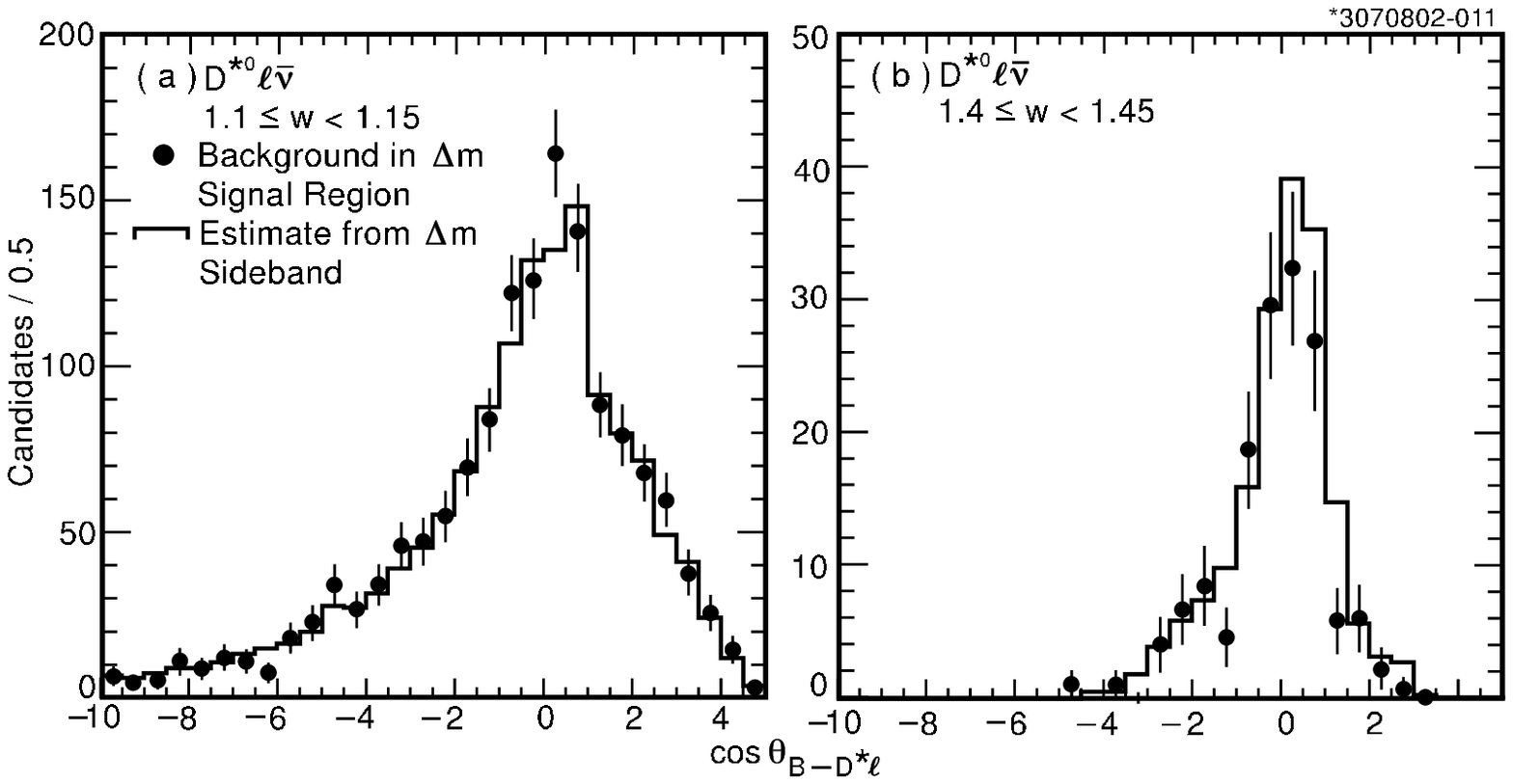,width=165mm}
\caption{\label{fig:comb_ds0}
From Monte Carlo, the $\cby$ distribution for $\dstzlnu$ 
combinatoric background candidates in the $\deltam$ signal region (points)
and for scaled candidates from the $\deltam$ sideband (histogram) for 
(a) $1.1 \le w < 1.15$ and (b) $1.4 \le w < 1.45$.  The sideband events
are normalized using a fit to the $\deltam$ distribution as described
in the text.}
\end{figure*}

As a test of this background estimate, we carry out the same procedure
used in data on a sample of 16 million simulated $\bbbar$ events.  Because
combinatoric background originates from random combinations of tracks
and showers, we expect our Monte Carlo, which is tuned to reproduce
track and shower multiplicity and momentum distributions of $B$
decays, to provide a reliable check of the background estimation
procedure.  We compare the true background in the $\deltam$ signal
region with the background estimate formed using the $\deltam$ sideband
region.  There is a concern that kinematic differences between
candidates in the $\deltam$ signal and sideband regions could cause a
difference in the $\cby$ shape of the estimated and true backgrounds.
Figures~\ref{fig:comb_dsp} and~\ref{fig:comb_ds0} show the true and
estimated backgrounds for the Monte Carlo sample.  We observe that the
shapes do differ for $\dstplnu$, consistent with the effect of the strong
momentum dependence of the slow-pion efficiency
(Section~\ref{sec:slowpi}).  The agreement is better for $\dstzlnu$.  We
evaluate the systematic error from this sideband technique in
Section~\ref{sec:comb_sys}. 

This method of background estimation overlooks a small component of the
combinatoric background, a component that arises from $D^*$ decays in
which the slow pion is properly found but the $D^0$ candidate is
constructed from the products of a $D^0$ decay other than $D^0\to
K^-\pi^+$.  Although the $D^0$ is misreconstructed,
this background will still peak in the $\deltam$ signal region.
Most of these candidates have $\mkpi$ below our signal
region, but Monte Carlo simulation shows that $D^0\to \rho^+ \pi^-/\rho^-
\pi^+$ decays could contribute a few candidates to the $\mkpi$ and $\deltam$ 
signal regions.  Although these decay modes have not yet been observed, a
combined branching fraction of 1.3\% is plausible given the measured 
branching fraction of all $D^0\to \pi^+ \pi^- \pi^0$ decays; with this
branching fraction, these modes would increase our $\dstlnu$ yield by
$(0.3 \pm 0.2)\%$. 
In a sample of 16 million simulated $\bbbar$ decays, 
several other modes also contribute, bringing the total contribution
to $(0.5 \pm 0.3)\%$.  The contributing modes are listed in
Table~\ref{tab:mdleak}.
As this contribution has little effect on our
results, and as the branching fractions of the main contributing modes
($D^0\to \rho^+ \pi^-$ and $D^0\to \rho^- \pi^+$) are unmeasured, we
account for it in the combinatoric background systematic error, but
otherwise neglect it.

\begin{table}
\caption{\label{tab:mdleak}
Decay modes of the $D$, other than $D^0\to K^-\pi^+$, that
are not fully subtracted by the $\deltam$ sideband.  The third column
shows the expected contribution (relative to $D^0\to K^-\pi^+$) to the
$\dstlnu$ yield from each mode in the $\mkpi$ and $\deltam$ signal regions.
}
\begin{ruledtabular}
\begin{tabular}{lcc}
Mode                                & {Branching Fraction(\%)}
	& Contribution(\%)   \\\hline
$D^0\to K^+ K^-$                    & $0.425\pm0.016$  \footnotemark[1]
	& $0.05\pm0.03$\\
$D^0\to \pi^+ \pi^-$                & $0.152\pm0.009$  \footnotemark[1]
	& $0.02\pm0.02$\\
$D^0\to K^- \bar{\ell}\nu$          & $3.47\pm0.17$    \footnotemark[1]
	& $0.02\pm0.02$\\  
$D^0\to \pi^+ \pi^- \pi^0$          & $1.6\pm1.1$      \footnotemark[1]\footnotemark[2]
	& $0.33\pm0.24$\\ 
$D^0\to K^- \rho^+$                 & $10.8\pm1.0$     \footnotemark[1]
	& $0.02\pm0.02$\\
$D^0\to K^{*-} \pi^+$               & $1.7\pm0.2$      \footnotemark[1]
	& $0.01\pm0.01$\\
$D^0\to \pi^- \bar{\ell}\nu$        & $0.37\pm0.06$    \footnotemark[1]\footnotemark[3]
	& $0.07\pm0.05$\\\hline
Total   &  
	& $0.52\pm0.25$ \\
\end{tabular}
\end{ruledtabular}
\footnotetext[1]{From Ref.~\onlinecite{pdg}.}
\footnotetext[2]{The simulation includes nonresonant 
$D^0\to \pi^+\pi^-\pi^0$ and resonant $D^0\to \rho\pi$ submodes.}
\footnotetext[3]{Assuming lepton universality, we use the
$D^0\to\pi^-e^+\nu$ branching fraction for $D^0\to\pi^-\bar{\ell}\nu$.}
\end{table}

\subsubsection{Uncorrelated background}
Uncorrelated background arises when the $D^*$ and lepton come from the
decays of different $B$ mesons in the same event.  This background
accounts for approximately 5\% of the candidates in the signal region for
both $\dstplnu$ and $\dstzlnu$ decays, contributing 6\% and 9\% relative
to $\dstplnu$ and $\dstzlnu$, respectively.
We obtain the $\cby$ distribution of this background by 
simulating each of the various sources of uncorrelated $D^*$'s and leptons and 
normalizing each one based on rates measured from or constrained by the
data. We classify the $D^*$ and the lepton according to their
respective sources because different sources give different momentum
spectra for the $D^*$ and lepton, and therefore different
distributions in $\cby$. 

There are three components of uncorrelated background that contribute
to both the $\dstplnu$ and $\dstzlnu$ modes.  The first component
consists of a lower-vertex $\dst$ ({i.e.}, from $b\to c$
transitions) combined with a secondary lepton ({i.e.}, from $b\to
c\to s\bar{\ell}\nu$) (primary leptons from the other $B$ have the wrong
charge correlation); this is the largest component for the $\dstplnu$
mode. Secondly, uncorrelated background can also occur when the $B^0$ and
$\bar{B}^0$ mix or when a $\bar{D^*}$ from the upper-vertex ({i.e.},
from $b\to \bar{c}$, as in $b\to c\bar{c}s$) is combined with a
primary lepton ({i.e.}, from $b\to c(u)\ell\bar{\nu}$). Finally, in the
$\dstzlnu$ case, the largest source of uncorrelated background consists
of candidates in which the $K$ and $\pi$ from a lower-vertex $D^*$ have
been exchanged and paired with a primary lepton from the other $B$.
(This background does not occur for $\dstplnu$ because we constrain the
charge of the slow-pion candidate to be opposite to that of the kaon.)

\begin{table}
\caption{\label{tab:uncorr_norm}
The rate per $\bbbar$ pair used to normalize the $\dst$ elements of the 
uncorrelated background.  The errors indicate the variation
used to assess the systematic uncertainty in the background.}
\begin{ruledtabular}
\begin{tabular}{ccc}
Rate & $p_{D^*}\le 1.3$~\gevc & $p_{D^*}>1.3$~\gevc \\\hline
lower-vertex, $\dstp$ & 0.281$\pm$0.032 & 0.242$\pm$0.015 \\
lower-vertex, $\dstz$ & 0.231$\pm$0.031 & 0.272$\pm$0.014 \\
upper-vertex, $\dstp$ & 0.048$\pm$0.024 & 0.012$\pm$0.006 \\
upper-vertex, $\dstz$ & 0.048$\pm$0.024 & 0.004$\pm$0.002 \\
\end{tabular}
\end{ruledtabular}
\end{table}

We first determine the production rate of upper-vertex $D^*$'s from
$B$ decays using the measured branching fractions of modes 
such as $\bar{B}\to D^{(*)}\bar{D}^{(*)}\bar{K}^{(*)}$\cite{DDK}.  We
do this in 
two $D^*$ momentum bins, relying on our simulation of such decays for
the $D^*$ momentum distribution. To determine the lower-vertex
$D^*$ production rate, we measure the 
rate of inclusive $D^*$ production from $B$ decays in the 
data in each momentum bin and subtract the upper-vertex
contribution from each. The results are shown in 
Table~\ref{tab:uncorr_norm}.
We determine the background contribution from $D$'s 
reconstructed with exchanged $K$'s and $\pi$'s by studying inclusive 
$\dstp$ decays with the charge correlation of the slow pion reversed. 
%The simulation reproduces the $K$--$\pi$ exchange background
%accurately, as expected since this effect is primarily kinematic.

We normalize the primary lepton decay rate for leptons with momenta between 
0.8~\gevc\ and 2.4~\gevc\ to its measured value of
$(8.99\pm0.42)$\%~\cite{roywang}, 
where the error includes statistical and systematic errors; since this 
measurement was made at CLEO, we include only the systematic errors that 
are uncorrelated with our analysis. Likewise, we adjust the secondary lepton 
rate for leptons with momenta between 0.8 \gevc\ and 2.4 \gevc\ to its
measured value of  
$(1.53\pm0.12)$\%~\cite{roywang}. Finally, we adjust $\chi_d$, the 
$B^0 - \bar{B}^0$ mixing probability, to its measured value of 
$0.174\pm0.009$~\cite{pdg}. 

\subsubsection{Correlated background}
Correlated background candidates are those in which the $D^*$ and
lepton are decay products of the same $B$, but the decay was not
$\bdstlnu$ or $\bdstxlnu$.  The most common sources are $\bar{B}\to
D^*\tau^-\bar{\nu}$ followed by leptonic $\tau$ decay, and 
$\bar{B} \to D^* D_s^{(*)-}$ followed by semileptonic  decay of the $D_s^-$.
The uncorrelated background contributes 0.5\% and 0.2\% compared to
the $\dstplnu$ and $\dstzlnu$ signals, respectively.  The background is
small; we therefore rely on our Monte Carlo simulation to quantify it. 
The decay modes and branching fractions used are listed in
Table~\ref{tab:corr_modes}.

\begin{table}
\caption{\label{tab:corr_modes}
Modes that contribute to the correlated background, their
assumed branching fractions (BF), and fraction of the total correlated
background.  The numbers given are for the $\dstzlnu$ mode. The 
contributions to the $\dstplnu$ mode are similar.}
\begin{ruledtabular}
\begin{tabular}{ldd} 
Mode& \multicolumn{1}{c}{BF~\cite{pdg}} (\%) & 
\multicolumn{1}{c}{Fraction(\%)} \\\hline
$\bar{B}\to\dstz X \tau^-\bar{\nu}$                &1.65  &  36.0   \\
$\bar{B}\to D_s^{*-}\dstz$                         &3.08  &  18.7   \\
$\bar{B}\to\dstz D^{(*)}K^{(*)}$                   &5.6   &  18.7   \\
$\bar{B}\to\dstz\bar{D^*}$                         &3.4   &  15.1   \\
$\bar{B}\to D_s^-\dstz$                            &1.39  &   6.5   \\
$\bar{B}\to\dstz \gamma X$; $\gamma \to e^+ e^-$   &\multicolumn{1}{c}{---}
							  &   3.6   \\
$\bar{B}\to\dstz \pi^-$; $\pi^-\to \mu^- \bar{\nu}$&0.47  &   1.4   \\
\end{tabular}
\end{ruledtabular}
\end{table}

\subsubsection{Fake lepton background}
Fake lepton background arises when a hadron is misidentified as a lepton 
and is then used in our reconstruction. Fake leptons make up 0.5\% of
candidates in the 
signal region for $\dstplnu$ and 0.2\% for $\dstzlnu$; relative to signal,
the background contributions are about 0.5\%.  To assess this
background we repeat the analysis, using hadrons in place of the
lepton candidates.  After subtracting continuum and combinatoric
backgrounds, we normalize the $\cby$ 
distributions with the probability for a hadron to fake
an electron or muon. 
We measure the momentum-dependent 
fake probability using kinematically identified samples of
hadrons: pions are identified using $K_S^0\rightarrow \pi^+\pi^-$
decays, kaons using $D^{*+}\rightarrow D^0\pi^+\rightarrow
K^-\pi^+\pi^+$, and protons from $\Lambda\rightarrow p\pi^-$.  The
fake probabilities are then weighted by species abundance in $B$
decays and the momentum spectrum of hadronic tracks in events with an
identified $D^{*+}$ to obtain an average fake rate of 0.035\% for a
hadronic track to fake an electron and 0.68\% to fake a muon.

\subsubsection{$\dstlnu$ and $\dstxlnu$ $\cby$ distributions}
\label{sec:signal_cosby}
The $\cby$ distributions of $\dstlnu$ and $\dstxlnu$ decays are obtained
from simulated $\bbbar$ events in which one of the $B$'s is required to
decay to $\dstlnu$ or $\dstxlnu$. Since the other $B$ in the event also
decays, the $\cby$ distributions can contain the same backgrounds
listed above.
Using generator-level information, we veto all background
sources except the combinatoric background, for which we perform the same
$\deltam$ sideband subtraction used in the data.  In the sideband
subtraction, we use the signal-region to sideband ratios obtained from 
$\deltam$ fits for 
the data. (Comparison of these ratios for data and simulated $\bbbar$
decays shows them to be compatible.)  This sideband subtraction correctly
accounts for the small number of signal decays that populate the
$\deltam$ sideband.

\subsection{$\bdstlnu$ yields}

\begin{figure*}
\epsfig{file=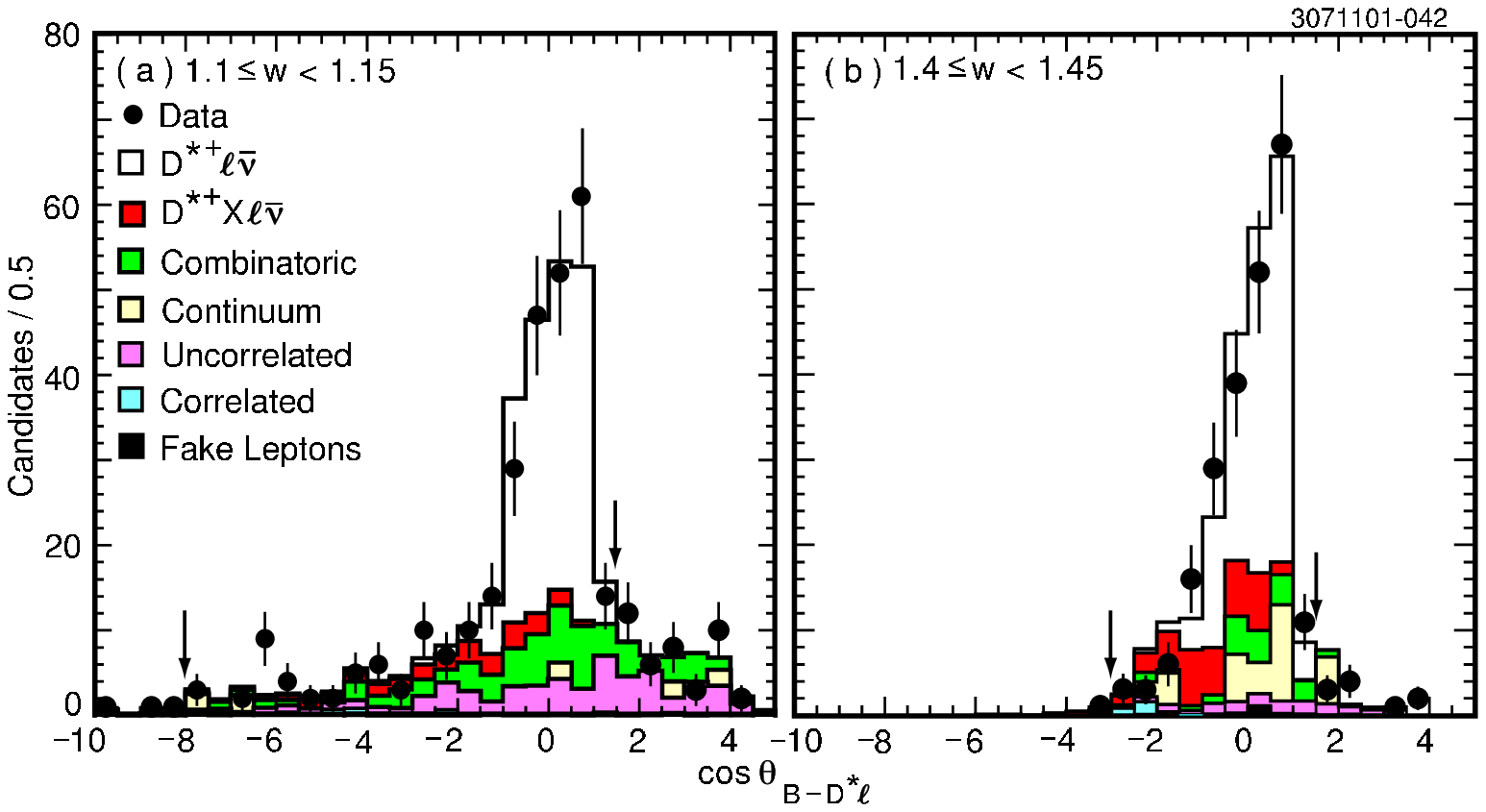,width=165mm}
\caption{\label{fig:dstpwbin39}
The $\cby$ distribution (solid circles) for $\dstplnu$ in 
the intervals (a) $1.1 \le w <1.15$ and (b) $1.4 \le w < 1.45$ with the
results of the fit superimposed (histogram). The arrows indicate the
fit ranges.}
\end{figure*}

\begin{figure*}
\epsfig{file=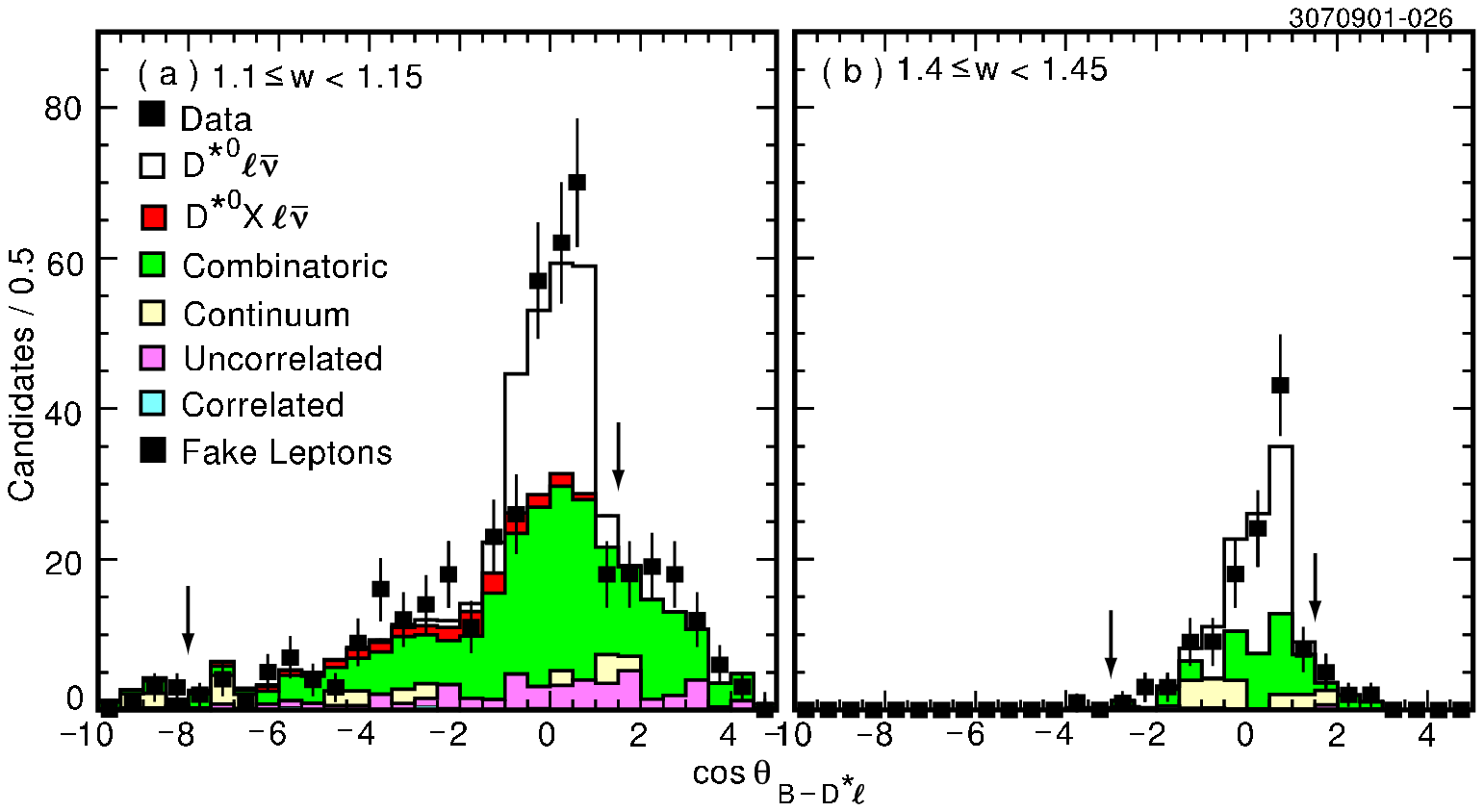,width=165mm}
\caption{\label{fig:dstzwbin39}
The $\cby$ distribution (solid squares) for $\dstzlnu$ in the
intervals (a) $1.1 \le w <1.15$ and (b) $1.4 \le w < 1.45$
with the results of the fit superimposed (histogram). The arrows
indicate the fit ranges.} 
\end{figure*}

\begin{figure*}
\epsfig{file=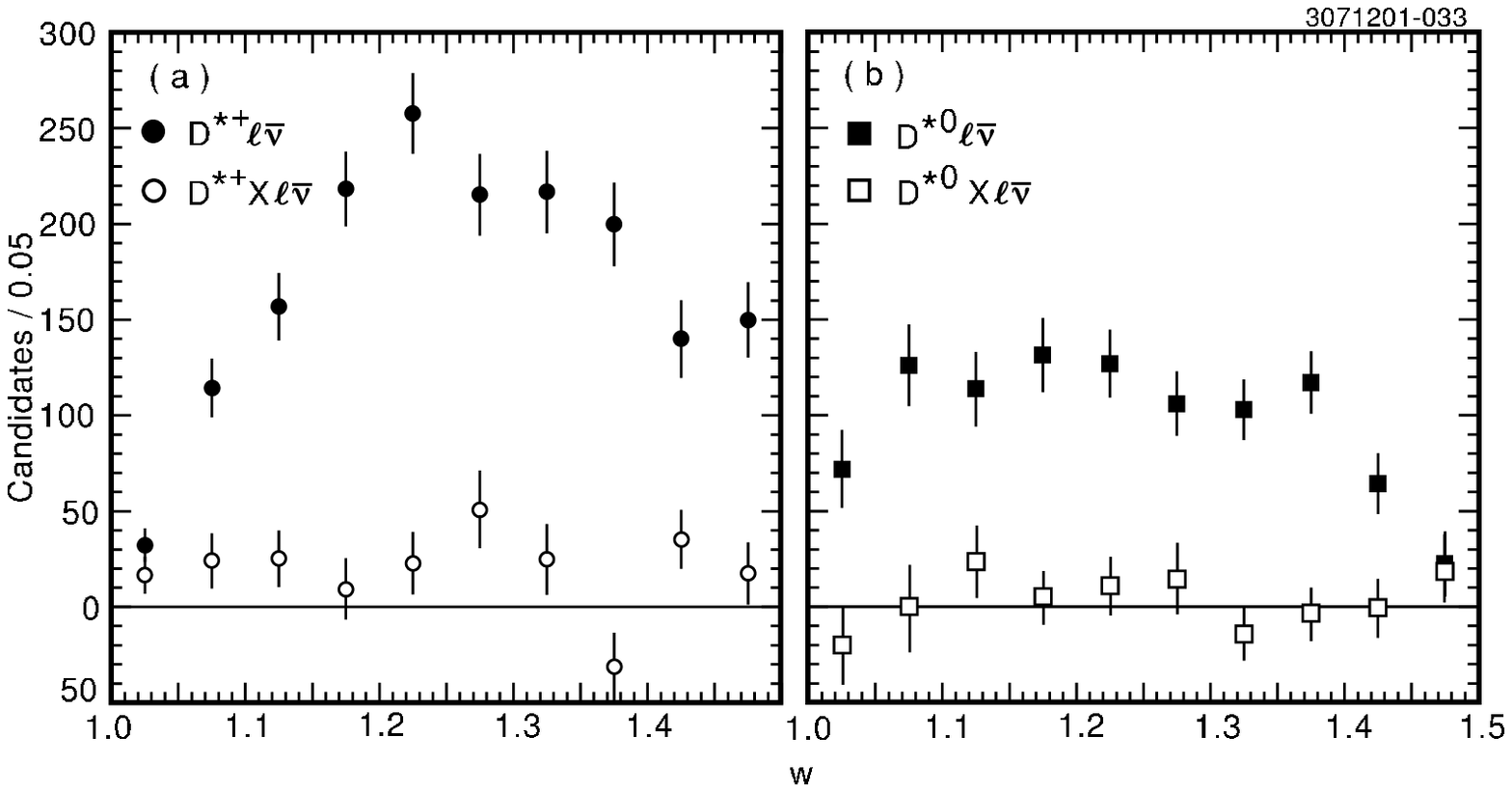,width=165mm}
\caption{\label{fig:yields}
The observed 
(a) $\dstplnu$ and $\dstp X \ell\bar{\nu}$ yields and 
(b) $\dstzlnu$ and $\dstz X \ell\bar{\nu}$ yields in each $w$ bin.}
\end{figure*}

Having obtained the distributions in $\cby$ of the signal and background
components, we fit for the yield of $\dstlnu$ candidates in each $w$ bin.
Two representative fits are shown for $\dstplnu$ in Fig.~\ref{fig:dstpwbin39}
and $\dstzlnu$ in Fig.~\ref{fig:dstzwbin39}.
The quality of the fits is good, as is agreement between the 
data and fit distributions outside the fitting region.  We summarize 
the observed $\dstlnu$ and $\dstxlnu$ yields in Fig.~\ref{fig:yields}.

The $\dstxlnu$ yields correspond to branching fractions of
${\cal B}(\bar{B}\to\dstpxlnu)= (0.97 \pm 0.24 \text{(stat.)})$\% and
${\cal B}(\bar{B}\to\dstzxlnu)= (0.32 \pm 0.55 \text{(stat.)})$\%.
These are somewhat lower than past
measurements~\cite{alephdstx,delphidstx}, but because the 
analysis is not optimized for these modes 
the systematic uncertainties on these branching fractions are large,
of order $\pm 30\%$ for $\dstp$ and $\pm 60\%$ for $\dstz$, dominated by
model dependence in the efficiency 
to satisfy our lepton momentum criteria, uncertainty in
the correlated and uncorrelated backgrounds, and radiative effects
in $\dstlnu$. 

\begin{figure*}
\epsfig{file=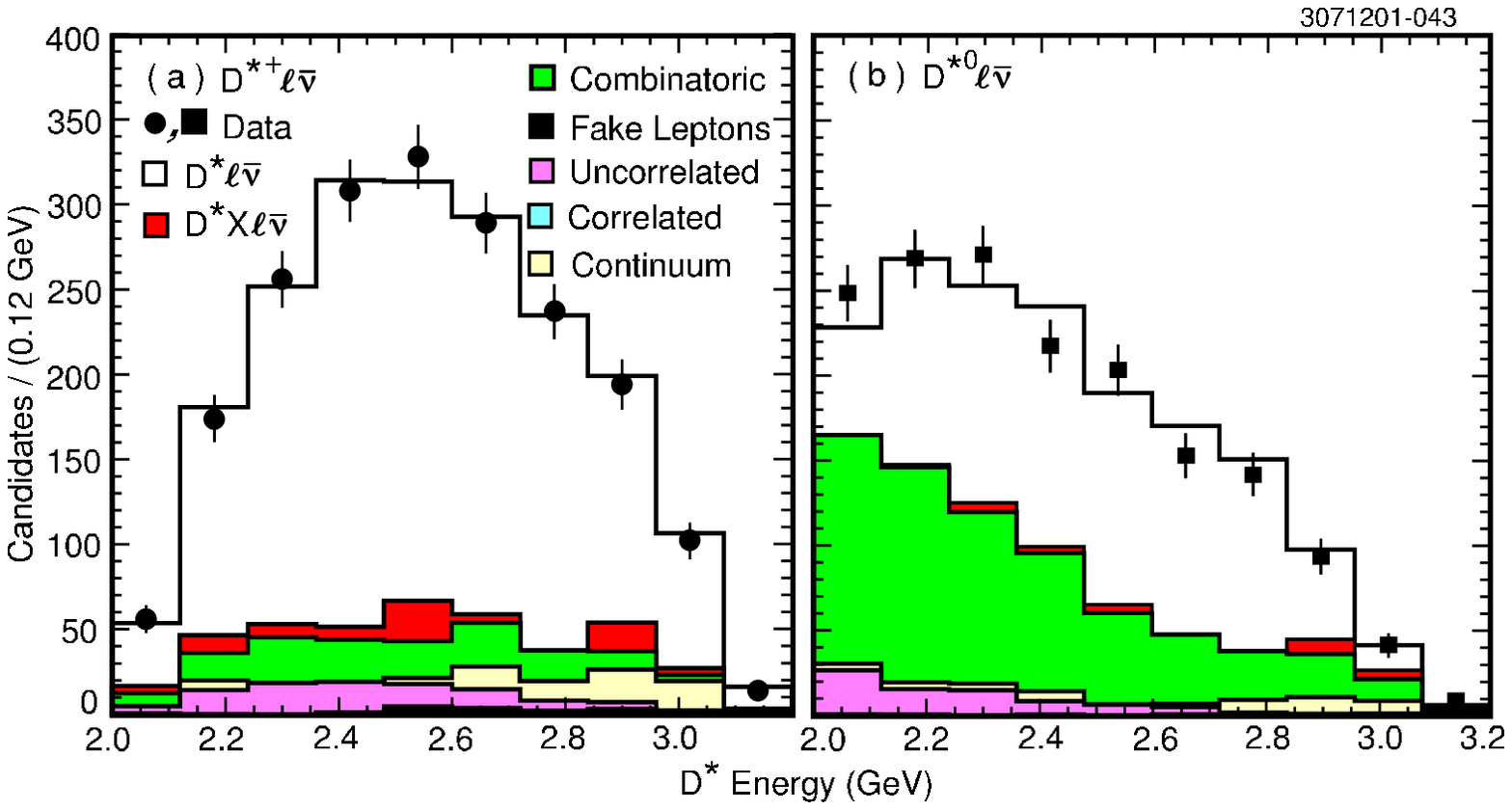,width=165mm}
\caption{\label{fig:edst}
The $D^*$ energy distribution of (a) $\dstplnu$ candidates and 
(b) $\dstzlnu$ candidates in the region
$\left\lvert\cby\right\rvert \le 1$ for all $w$ bins combined.} 
\end{figure*}

\begin{figure*}
\epsfig{file=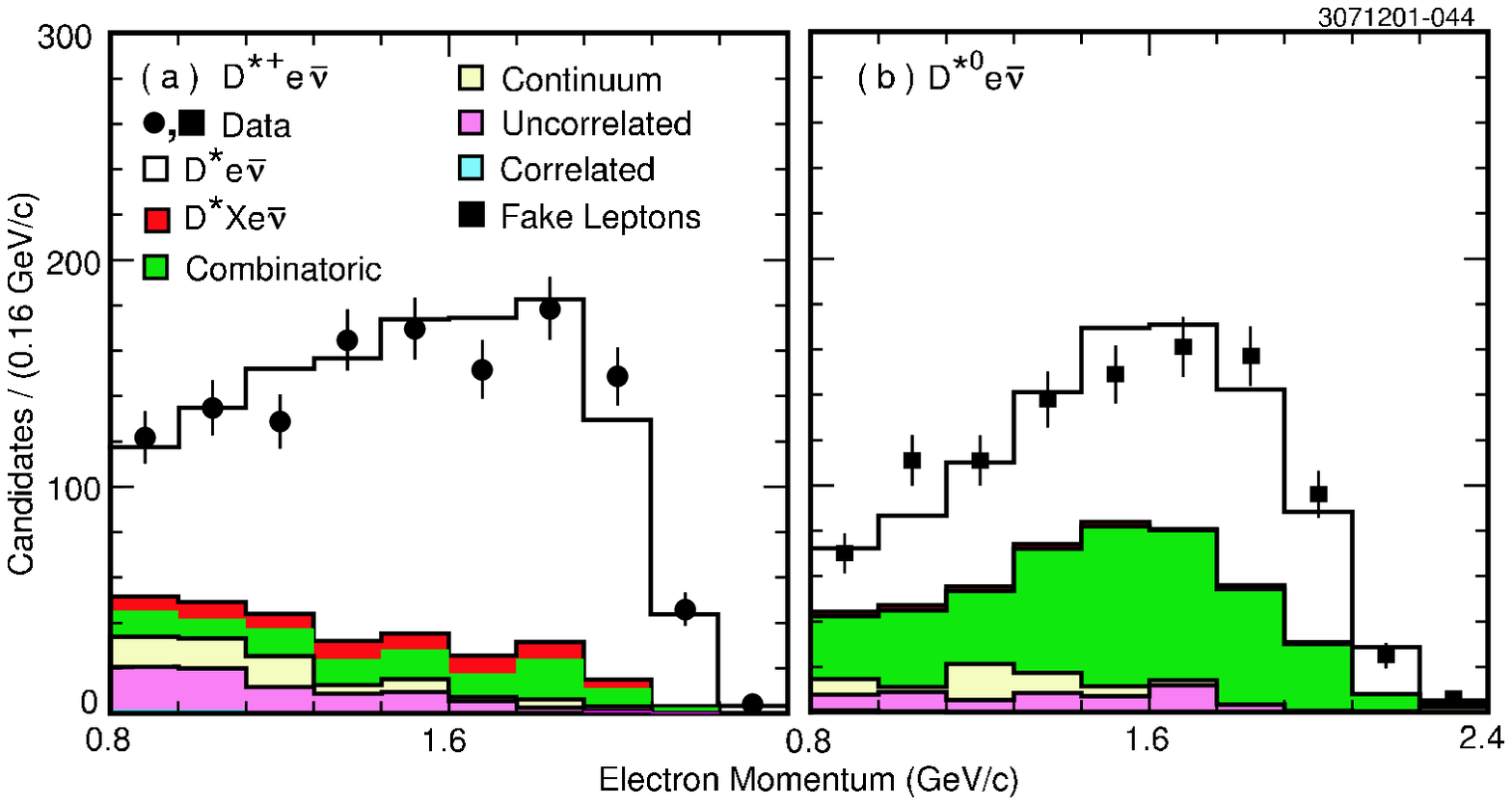,width=165mm}
\caption{\label{fig:elep}
The electron momentum spectrum for (a) $D^{*+}e^-\bar{\nu}$
candidates and 
(b) $D^{*0}e^-\bar{\nu}$ candidates in the region 
$\left\lvert\cby\right\rvert\le 1$ for all $w$ bins combined.
} 
\end{figure*}

\begin{figure*}
\epsfig{file=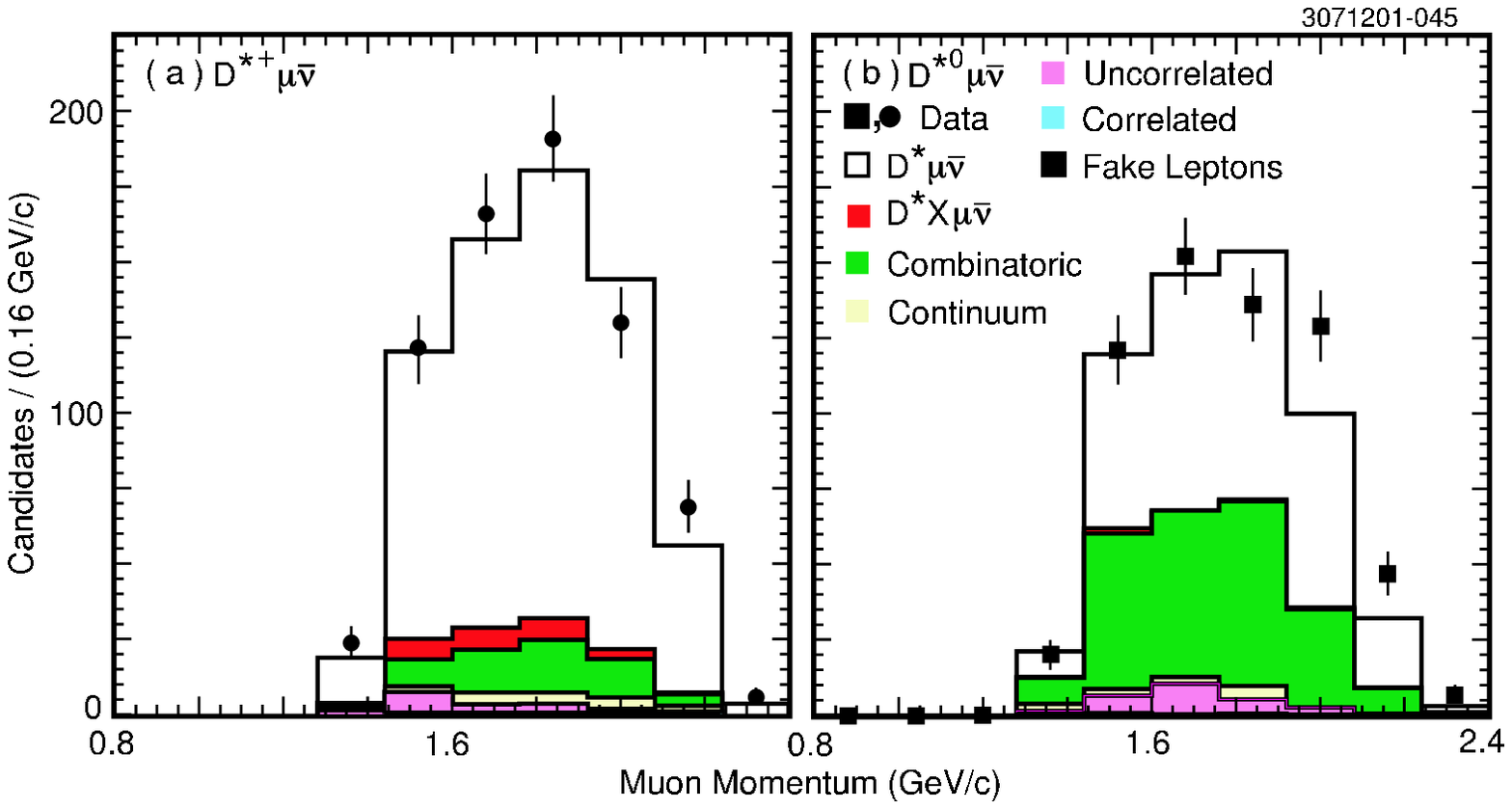,width=165mm}
\caption{\label{fig:mulep}
The muon momentum spectrum for (a) $D^{*+}\mu^-\bar{\nu}$ candidates and 
(b) $D^{*0}\mu^-\bar{\nu}$ candidates in the region 
$\left\lvert\cby\right\rvert\le 1$ for all $w$ bins combined.
} 
\end{figure*}

In order to test the quality of our $\cby$ fits and the modeling of the signal
and backgrounds, we compare the observed $D^*$ energy and lepton momentum 
spectra with expectations for candidates in the signal-rich region 
$\left\lvert\cby\right\rvert \le 1$.
The $\cby$ fits provide the normalizations of the $\dstlnu$ and
$\dstxlnu$ components. 
Figure~\ref{fig:edst} shows the $D^*$ energy distributions, and
Figs.~\ref{fig:elep} and~\ref{fig:mulep} show  
the electron and muon momentum spectra, repsectively, for $\dstplnu$
and $\dstzlnu$ candidates.  We find good agreement between the data
and our expectations.

\section{The $|V_{\lowercase{cb}}|$ Fit}
\label{sec:thefit}

The partial width for $\bdstlnu$ decays is given by~\cite{richburch} as
\begin{eqnarray}
\frac{d\Gamma}{dw}&=&\frac{G_F^2}{48 \pi^3}
(m_B-m_{D^*})^2 m_{D^*}^3 \sqrt{w^2-1}(w+1)^2  \\*
&&\times \left[1 + \left(\frac{4w}{w+1}\right)
\left(\frac{1-2wr+r^2}{(1-r)^2}\right)\right]
\vcb^2{\cal F}^2(w), \nonumber
\end{eqnarray}
where $m_B$ and $m_{D^*}$ are the $B$- and $D^*$-meson masses,
$r=m_{D^*}/m_B$, and the form factor $\fw$ is given by
\begin{equation}
\fw = 
\sqrt{\frac{\tilde{H}^2_0 + \tilde{H}^2_+ + \tilde{H}^2_-}
{1 + \left(\frac{4w}{w+1}\right)\left(\frac{1-2wr+r^2}{(1-r)^2}\right)}}
h_{A_1}(w).
\end{equation}
The $\tilde{H}_i$ are the helicity form factors
and are given by
\begin{eqnarray}
\tilde{H}_0(w) &=& 1 + \frac{w-1}{1-r}\left[1-R_2(w)\right],
\ {\text{and}} \\
\tilde{H}_\pm(w) &=& \frac{\sqrt{1-2wr+r^2}}{1-r}\left[
1\mp\sqrt{\frac{w-1}{w+1}}R_1(w)\right].
\end{eqnarray}
The form factor $h_{A_1}(w)$ and the form-factor ratios
$R_1(w) = h_V(w)/h_{A_1}(w)$ and $R_2(w) =[h_{A_3}(w)+rh_{A_2}(w)]/h_{A_1}(w)$
have been studied both experimentally and theoretically.  A CLEO 
analysis~\cite{ffprl} measured these form-factor parameters under the
assumptions 
that $h_{A_1}(w)$ is a linear function of $w$ and that $R_1$ and
$R_2$ are independent of $w$.  CLEO found 
\begin{eqnarray}
\frac{-1}{h_{A_1}(1)}\left.\frac{dh_{A_1}}{dw}\right\rvert_{w=1}
\equiv \rho^2 &=& 0.91 \pm 0.15 \pm 0.06, \nonumber \\
R_1 &=& 1.18 \pm 0.30 \pm 0.12, {\text{\ and}}  \nonumber \\
R_2 &=& 0.71 \pm 0.22 \pm 0.07, \nonumber
\end{eqnarray}
with the correlation coefficients $C(\rho^2,R_1)=0.60$, 
$C(\rho^2,R_2) = -0.80$, and $C(R_1, R_2) =-0.82$.  

$R_1(1)$ and $R_2(1)$ have been computed 
using QCD sum rules
with the results
$R_1(1)=1.27$ and $R_2(1)=0.8$ and estimated errors of
$0.1$ and $0.2$, respectively~\cite{neubert}, in good agreement
with the (later) experimental results. $R_1(w)$ and $R_2(w)$ 
are expected to vary weakly with $w$.
Most importantly for this analysis,
$\fone(=h_{A_1}(1))$ is relatively well-known
theoretically~\cite{dslnu_neubert,dslnu_falk},
thereby allowing us to disentangle it from $\vcb$.

Recently, dispersion relations have been used to constrain the
shapes of the form factors~\cite{lebed,caprini}. Rather than
expand the form factor in $w$, these analyses
expand in the variable $z=(\sqrt{w+1}-\sqrt{2})/(\sqrt{w+1}+\sqrt{2})$. 
The authors of Ref.~\onlinecite{caprini} obtain
\begin{widetext}
\begin{eqnarray}
\label{eq:formfact1}
h_{A_1}(w) &=& h_{A_1}(1)[1 - 8\rho^2z +
              ( 53\rho^2-15 )z^2 -
              ( 231\rho^2 - 91 )z^3], \\
\label{eq:formfact2}
R_1(w) &=& R_1(1) - 0.12(w-1) + 0.05(w-1)^2,\ {\text{and}}\\
R_2(w) &=& R_2(1) + 0.11(w-1) - 0.06(w-1)^2.
\label{eq:formfact3}
\end{eqnarray}
\end{widetext}

In our analysis, we assume that the form factor has the functional form 
given in Eqs.~\ref{eq:formfact1}-~\ref{eq:formfact3}.
We fit our yields as a function 
of $w$ for $\vcbf$ and $\rho^2$, keeping $R_1(1)$ and $R_2(1)$ fixed at 
their measured values. Our fit minimizes 
\begin{equation}
\chi^2 = \sum_{i=1}^{10}
\frac{[N_i^{obs} - \sum_{j=1}^{10}\epsilon_{ij}N_j]^2}
{\sigma_{N_i^{obs}}^2}, \label{eq:chi2}
\end{equation}
where $N_i^{obs}$ is the yield in the $i^{{\text{th}}}$ $w$ bin,
$N_j$ is the number of decays in the $j^{{\text{th}}}$ $w$ bin, and 
the matrix $\epsilon$ accounts for the reconstruction efficiency
and the smearing in $w$.

The efficiency matrix $\epsilon$ is calculated using simulated
$\dstlnu$ decays. A matrix element $\epsilon_{ij}$ represents the fraction of 
$\dstlnu$ decays generated in the $j$th $w$ bin
that are reconstructed in the $i$th $w$ bin. To be consistent with our method
for finding the $\cby$ distribution of $\dstlnu$ decays, described in
Section~\ref{sec:signal_cosby}, we subtract the combinatoric background
in the simulated decays using the $\deltam$ sideband and the data 
normalizations. 
We veto all other backgrounds
using generator-level knowledge of the simulated events. A single
element of the efficiency matrix is thus calculated using
\begin{equation}
\epsilon_{ij} = (S_{i}^{sig} - n_i S_{i}^{side})/S_{j},
\end{equation}
where $S_{i}^{sig}$ and $S_{i}^{side}$ are the number of non-vetoed candidates
reconstructed in the $i$th $w$ bin in the $\deltam$ signal and sideband 
regions, respectively, $n_i$ is the normalization of the $\deltam$ sideband 
region, and
$S_{j}$ is the number of $\dstlnu$ decays generated in the $j$th $w$ bin.

The efficiency matrix is nearly diagonal because the resolution in $w$
is about half the bin size.   The off-diagonal elements are only
appreciable for $|i-j|\le 1$.  The resolution becomes worse for larger
$w$ (see Fig.~\ref{fig:wrec}).
The efficiency matrix depends not only on the experimental selection
criteria but also on the form factor.  For the cuts described in this
paper and using the form factor described above, the diagonal elements
of the efficiency matrix vary from 4--14\% for $\dstplnu$ and from
5--11\% for $\dstzlnu$.
Although we bin in $w$, the efficiency matrix has a weak dependence on
the slope parameter $\rho^2$.  We iterate the fit, reevaluating the
efficiency matrix for the best-fit value of $\rho^2$.  A single
iteration is sufficient for convergence.

In Equation~\ref{eq:chi2}, $N_j$ is given by
\begin{equation}
N_j = 4 f_{00} N_{\Ufs}{\cal B}_{\dstp}
{\cal B}_{D^0} \tau_{B^0} \int_{w_j}dw \frac{d\Gamma}{dw}
\end{equation}
for $\dstplnu$, where $\tau_{B^0}$ is the $B^0$ lifetime~\cite{pdg}, 
${\cal B}_{\dstp}$ is the $\dstp \to D^0\pi^+$ branching
fraction~\cite{pdg},
${\cal B}_{D^0}$ is the $D^0 \to K^-\pi^+$ branching fraction,
$N_{\Ufs}$ is the number of $\Ufs$ events in the sample, 
and $f_{00}$ represents the $\Ufs\to B^0 \bar{B}^0$ branching
fraction.  For $\dstzlnu$,
\begin{equation}
N_j = 4 f_{+-} N_{\Ufs}{\cal B}_{\dstz}
{\cal B}_{D^0} {\cal B}_{\pi^0} \tau_{B^+}\int_{w_j}dw \frac{d\Gamma}{dw},
\end{equation}
where ${\cal B}_{\dstz}$ is the $\dstz\to D^0\pi^0$ 
branching fraction~\cite{pdg}, 
${\cal B}_{\pi^0}$ is the $\pi^0\to \gamma\gamma$ branching
fraction~\cite{pdg}, and $f_{+-}$ represents the  
$\Ufs\to B^+ B^-$ branching fraction.  
The values that we
use for the $B$ lifetimes and the various branching fractions are
listed in Table~\ref{tab:bfs}. For ${\cal B}(D^0\to K^-\pi^+)$,
we average the CLEO~\cite{cleo_kpi} and ALEPH~\cite{aleph_kpi} results 
after correcting
the former for final-state radiation (about a 2\% correction) to obtain
a branching fraction for the sum of radiative
and non-radiative decays.  We exclude the other results included
in the PDG average because they do not specify their treatment of
radiation.

\begin{table}
\caption{\label{tab:bfs}
The lifetimes and the branching fractions used in the
$\vcb$ fit. }
\begin{ruledtabular}
\begin{tabular}{lc}
$\tau_{B^+}$                      & (1.653$\pm$0.028) ps \\
$\tau_{B^0}$                      & (1.548$\pm$0.032) ps \\
${\cal B}(\dstp\to D^0\pi^+)$     & (67.7$\pm$0.5)\% \\
${\cal B}(\dstz\to D^0\pi^0)$     & (61.9$\pm$2.9)\% \\
${\cal B}(D^0\to K^-\pi^+)+{\cal B}(D^0\to K^-\pi^+\gamma)$ & (3.89$\pm$0.11)\% \\
${\cal B}(\pi^0\to \gamma\gamma)$ & (98.798$\pm$0.032)\% \\
\end{tabular}
\end{ruledtabular}
\end{table}

\begin{figure*}
\epsfig{file=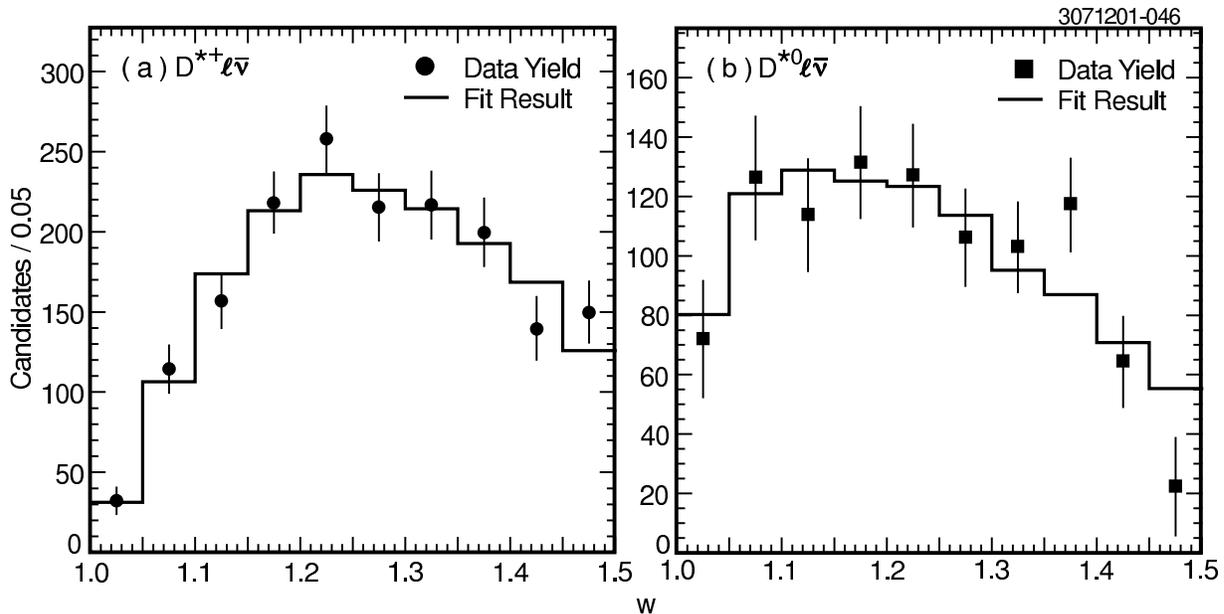,width=165mm}
\caption{\label{fig:fffit}
The results of (a) the $\dstplnu$ fit and (b) the $\dstzlnu$ fit.
The circles and squares are the data and the histogram shows the results
of the fit, done separately for each mode.}
\end{figure*}

We first fit $\dstplnu$ and $\dstzlnu$ separately, allowing as free
parameters $\vcbf$, $\rho^2$ and $f_{+-}$, with the last
of these constrained by adding a term 
$(R-R_0)^2/\sigma_R^2$ to the $\chi^2$ of Eq.~\ref{eq:chi2}.  
Here the double ratio $R\equiv[f_{+-}/(1-f_{+-})](\tau_{B^+}/\tau_{B^0})$
is compared to a measurement of the same double ratio ($R_0\pm\sigma_R$)
in Ref.~\onlinecite{sylvia},
and we have explicitly assumed $f_{00}+f_{+-}=1$.
The results of the 
separate fits are shown in Fig.~\ref{fig:fffit}. For $\dstplnu$, we find
\begin{eqnarray}
\vcbf &=& 0.0424 \pm 0.0018, \nonumber \\
\rho^2 &=& 1.60 \pm 0.11, \ {\text{and}} \nonumber \\
\chi^2 &=& 6.6/8 \,{\text{degrees\ of\ freedom\ (dof.)}}. \nonumber
\end{eqnarray}
These parameters imply $\Gamma = 0.0380\pm 0.0019$ \invps.
For $\dstzlnu$, we find
\begin{eqnarray}
\vcbf &=& 0.0436 \pm 0.0026, \nonumber\\
\rho^2 &=& 1.56 \pm 0.18,\ {\text{and}} \nonumber\\
\chi^2 &=& 9.5/8 \,{\text{dof.}} \nonumber
\end{eqnarray}
These parameters imply $\Gamma = 0.0415\pm0.0027$ \invps.
The results from $\dstplnu$ and $\dstzlnu$ are consistent with each other.

\begin{figure}
\epsfig{file=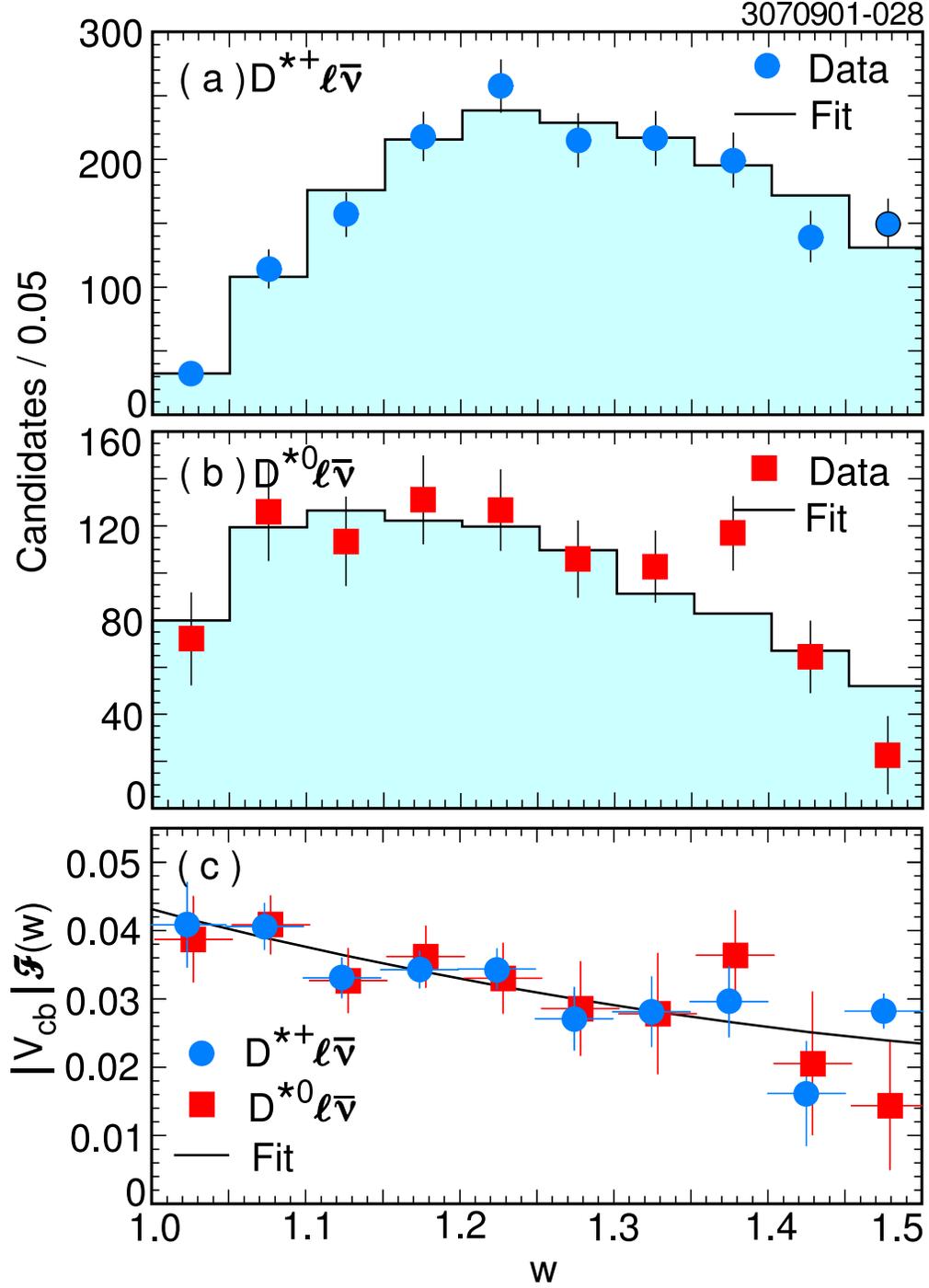,width=135mm}
\caption{\label{fig:combinedfit}
The results of the combined fit to the $w$ distribution:
(a) the $\dstplnu$ yields (circles) with the results
of the fit superimposed (histogram) and 
(b) the $\dstzlnu$ yields (squares) and fit (histogram).
In (c) the curve shows the best-fit $\vcb\fw$, and the circles
(squares) are the $\dstplnu$ ($\dstzlnu$) yields corrected for
efficiency, smearing, and all terms in the differential decay rate
apart from $\vcb\fw$.}
\end{figure}

We also do a combined fit to the $\dstplnu$ and $\dstzlnu$ data. In
minimizing, $\chi^2$ is the sum of the separate $\dstplnu$ and
$\dstzlnu$ $\chi^2$'s, but including the term constraining
$(f_{+-}/f_{00})(\tau_{B^+}/\tau_{B^0})$ only once.
The results of the fit are displayed in Fig.~\ref{fig:combinedfit}, and the 
parameter values are:
\begin{eqnarray}
\vcbf    &=&    0.0431 \pm    0.0013,  \nonumber \\
f_{+-}           &=&    0.521  \pm    0.012, \ {\text{and}}  \nonumber\\
\rho^2           &=&    1.61   \pm    0.09,\ {\text{with}}  \nonumber \\
\chi^2           &=&    16.8 / 18 \,{\text{dof.}}  \nonumber
\end{eqnarray}
These parameters give $\Gamma = 0.0394\pm0.0012$ \invps.
Not surprisingly, the values of $\vcbf$ and $\rho^2$ are strongly
correlated.  The correlation coefficients are $C(\vcbf,\rho^2) = 0.865$,
$C(\vcbf, f_{+-})=0.130$, and $C(\rho^2,f_{+-})=-0.075$.

When we remove the constraint on $f_{+-}$ in the fit, we find
$f_{+-}=0.532 \pm 0.016$.  This is in agreement with the recent
CLEO measurement~\cite{sylvia}, which implies
$f_{+-}=0.510 \pm 0.017 ^{+0.011}_{-0.010}$.

These results may be compared to our previous analysis~\cite{oldcleo}
that analyzed a subset (approximately 50\%) of the current data
finding a smaller value for $\vcbf$.  The increase may be
attributed to several effects.  Changes in the measured values of
the $D^0$ and $D^*$ 
branching fractions and $B$ lifetimes cause a 2.3\% increase.
Inclusion of final-state radiation shifts $\vcbf$ by 2.4\%.  More
significantly, use of the improved form factor gives an increase of
5.7\%.  The new form factor has positive curvature, which results in an
increase when extrapolating to $w=1$.  The analysis of Caprini
\etal~\cite{caprini} shows there is correlation between the curvature 
and slope of the form factor, making this effect more pronounced for
the large slope preferred by our data.  The remainder of the increase
results from the larger data sample.

We test the compatibility of the old and new analyses by restricting 
the new analysis to the same subset of data and fitting using the old
form factor.\footnote{
Specifically, in \cite{oldcleo} we used a linear form factor 
$h_{A_1}(w) = h_{A_1}(1) \left[1-\rho^2_{A_1}(w-1)\right]$, 
and $R_1=R_2=1.$}
Adjusting for common values for $D^0$ and $D^*$ branching fractions
and $B$ lifetimes (Table~\ref{tab:bfs}) we find a change in $\vcbf$ of
$0.0020 \pm 0.0010 \pm 0.0022$, where the first error is statistical
(assessed conservatively assuming all candidates in the old analysis
are found in the new analysis) and the second error is an estimate of the
uncorrelated systematic uncertainties.  The largest of the latter are
due to slow pion efficiency, taken to be uncorrelated because of
significant differences in the tracking algorithms used in the two
analyses.  We conclude the old and new analyses are compatible within
the systematic uncertainties.  Because our new analysis includes the
data reported previously and takes advantage of theoretical
improvements in the form factor, the results reported here supersede
our previous results.

\section{Systematic Uncertainties}
The systematic uncertainties are summarized in Table~\ref{tab:syserr_all}.
The dominant systematic uncertainties arise from our background
estimations and from our knowledge of the slow pion reconstruction efficiency.

\begin{table*}
\caption{\label{tab:syserr_all}
The fractional systematic uncertainties, given in percent, for the
$\bdstplnu$ fit, the $\bdstzlnu$ fit, and the combined fit.}
\begin{ruledtabular}
\begin{tabular}{lddddddddd} 
        & \multicolumn{3}{c}{$\dstplnu$ fit} & 
          \multicolumn{3}{c}{$\dstzlnu$ fit} & 
          \multicolumn{3}{c}{Combined fit} \\
Source  
&\multicolumn{1}{c}{$\vcbf$}&\multicolumn{1}{c}{$\rho^2$}&\multicolumn{1}{c}{$\Gamma$}  
&\multicolumn{1}{c}{$\vcbf$}&\multicolumn{1}{c}{$\rho^2$}&\multicolumn{1}{c}{$\Gamma$}  
&\multicolumn{1}{c}{$\vcbf$}&\multicolumn{1}{c}{$\rho^2$}&\multicolumn{1}{c}{$\Gamma$}  
 \\ \hline

Backgrounds                     & 1.8 & 3.0 & 1.8  
                                & 2.4 & 5.0 & 2.2 
				& 1.8 & 3.1 & 1.7 \\
Reconstruction efficiency       & 4.4 & 5.0 & 4.9  
				& 3.5 & 6.2 & 6.5 
				& 2.9 & 3.2 & 4.6 \\
$B$ momentum \& mass            & 0.2 & 0.0 & 0.5  
				& 0.6 & 0.5 & 0.8 
				& 0.1 & 0.1 & 0.2 \\
$\bdstxlnu$ model               & 0.3 & 3.5 & 1.2  
				& 1.2 & 2.7 & 0.5  
				& 0.3 & 1.6 & 0.9 \\ 
Final-state radiation           & 0.7 & 0.3 & 1.1  
				& 0.8 & 0.5 & 1.2  
				& 0.7 & 0.3 & 1.1 \\ 
Number of $B\bar{B}$ events     & 0.9 & 0.0 & 1.8  
				& 0.9 & 0.0 & 1.8 
				& 0.9 & 0.0 & 1.8 \\
\hline
Subtotal                        & 4.9 & 6.8 & 5.8  
				& 4.6 & 8.5 & 7.3 
				& 3.6 & 4.8 & 5.4 \\
\hline
$\tau_B$ and branching fractions& 1.5 & 0.0 & 3.0  
				& 2.8 & 0.0 & 5.6 
				& 1.8 & 0.0 & 3.5 \\
$R_1(1)$ and $R_2(1)$           & 1.6 &11.7 & 1.8  
				& 1.1 &14.3 & 1.8 
				& 1.4 &12.0 & 1.8 \\
\hline
Subtotal                        & 2.2 &11.7 & 3.5  
				& 3.0 &14.3 & 5.9 
				& 2.3 &12.0 & 3.9 \\
\hline
Total                           & 5.3 &13.5 & 6.8  
				& 5.5 &16.6 & 9.3  
				& 4.3 &13.0 & 6.6 \\ 
\end{tabular}
\end{ruledtabular}
\end{table*}

\subsection{Background uncertainties}
Here we present the systematic uncertainties from our backgrounds.

\subsubsection{Continuum background}
Our estimate of background from $e^+e^-\to q{\bar q}$ is taken from
data collected below the $\Ufs$.  The shape of this background in each
$w$ bin is taken from off-resonance data, where we scale the energy
of the $D^*$ and lepton to reflect the difference in the on- and
off-resonance center-of-mass energies.  This scaling applies to
computation of $w$ and $\cby$.  The resulting distribution is scaled by
the ratio $({\cal L}_{\text{on}}/{\cal L}_{\text{off}})
(E_{\text{off}}^2/E_{\text{on}}^2)$, where the first factor is the
ratio of on- to off-resonance 
luminosities and the $E^2$ ratio corrects for the $1/s$ dependence of
the hadronic cross section.

The uncertainty on the normalization is small and has negligible effect
on the results because the continuum background itself is small.
To assess the systematic uncertainty from
the $D^*$ and lepton energy scaling, we compare our results with the
scaling to those obtained without it.
The systematic uncertainties are taken to be half this difference, and
are 0.03\%, 0.2\%, and 0.1\% 
for $\vcbf$, $\rho^2$, and $\Gamma(\bdstlnu)$, respectively.

\subsubsection{Combinatoric background}
\label{sec:comb_sys}
Our method for combinatoric background subtraction assumes that the
$\cby$ distribution of candidates in the $\deltam$ sideband matches that of
those in the signal region.  The Monte Carlo simulation should
reproduce any differences well since they arise from kinematic
effects.  We use a sample of 16 million Monte Carlo-simulated inclusive
$\bbbar$ events to test this assumption.  (See also
Figs.~\ref{fig:comb_dsp} and \ref{fig:comb_ds0} and discussion in
Section~\ref{sec:comb}.) 
We perform our analysis 
on the simulated events twice, once using the combinatoric background
subtraction procedure outlined in Section~\ref{sec:comb} and once 
using the absolutely normalized ``true'' combinatoric background,
{i.e.}, $D^*$ candidates in the $\deltam$ signal region that do not
arise from the decay of a $D^*$, in 
place of the $\deltam$ sideband distribution.  
Use of the ``true'' background instead of the estimate results in a
shift in the combined fit of $(-1.3\pm0.9)$\% in $\vcbf$,
$(-2.6\pm1.3)$\% in $\rho^2$, and $(-0.6 \pm 0.9)$\% in $\Gamma$.  
Any bias from the use of the $\deltam$ sideband to estimate the
combinatoric background is smaller than the statistical uncertainty of
the fit to the data. 
We conservatively assign systematic errors equal to the quadrature sum of
the shift and its statistical uncertainty, a total of 1.6\% for $\vcbf$.

The normalization of the background relies on the fits to the $\deltam$ 
distributions.  We assign an uncertainty for this by repeating our
analysis with different functional forms used to fit $\deltam$. We also
include a 0.1\% uncertainty because the simulated $\deltam$ signal
peaks are shifted a few tenths of an MeV lower than the data.  The
statistical error on the background normalization is included in the
statistical errors on our result.

The final contribution to the systematic uncertainty from our combinatoric
background estimate comes from the decay modes other than $D\to K\pi$ 
that are reconstructed in our $\mkpi$ signal region. The specific modes
were given in Table~\ref{tab:mdleak}. We find the total contribution to
our $\dstlnu$ yield from this source is (0.5$\pm$0.3)\%.
We add the yield and its uncertainty in quadrature to get a 0.6\% uncertainty
on our $\dstlnu$ yield. Because $\vcbf$ is proportional to the amplitude
rather than the rate, its error is half as big. 
We find the total error on $\vcbf$ due to the combinatoric background to be 
1.6\%. The errors from all components are summarized in 
Table~\ref{tab:combsys}.

\begin{table*}
\caption{\label{tab:combsys}
The uncertainties due to the combinatoric background.}
\begin{ruledtabular}
\begin{tabular}{lccc} 
Variation  & $\vcbf$ (\%)&$\rho^2$(\%) & $\Gamma(\bdstlnu)$(\%) \\\hline
$\dstplnu$ fit &&&\\
\hspace{5mm}$\cby$ distribution of sideband candidates & 1.6  & 2.7 & 1.3 \\
\hspace{5mm}$\deltam$ sideband normalizations          & 0.2  & 0.3 & 0.3 \\
\hspace{5mm}Non-$K\pi$ decays(see Table~\ref{tab:mdleak})  & 0.3  & 0.0 & 0.6 \\\hline
Total                                             & 1.6  & 2.7 & 1.4 \\\hline
$\dstzlnu$ fit &&&\\
\hspace{5mm}$\cby$ distribution of sideband candidates & 2.2  & 4.8 & 1.5 \\
\hspace{5mm}$\deltam$ sideband normalizations          & 0.3  & 0.6 & 1.2 \\
\hspace{5mm}Non-$K\pi$ decays(see Table~\ref{tab:mdleak}) & 0.3  & 0.0 & 0.6 \\\hline
Total                                             & 2.2  & 4.9 & 2.0 \\\hline
combined fit &&&\\
\hspace{5mm}$\cby$ distribution of sideband candidates & 1.6  & 2.9 & 1.1 \\
\hspace{5mm}$\deltam$ sideband normalizations         & 0.2  & 0.3 & 0.6 \\
\hspace{5mm}Non-$K\pi$ decays(see Table~\ref{tab:mdleak}) & 0.3  & 0.0 & 0.6 \\\hline
Total                                             & 1.6  & 2.9 & 1.3 \\\hline
\end{tabular}
\end{ruledtabular}
\end{table*}

\subsubsection{Uncorrelated background}
The main source of uncertainty from the uncorrelated background
is the normalization of the various contributions.  Of these, 
the most important is the normalization of the upper-vertex $D^*$
decays, which we vary by 50\%. Smaller uncertainties arise from the primary and 
secondary lepton rates, the uncertainty in $B^0-\bar{B}^0$ mixing, and the
uncertainty in the rate of exchanging $K$ and $\pi$ particles in $D^{*0}$
candidates. The effects of varying these rates are summarized in 
Table~\ref{tab:uncorrsys}.  The systematic uncertainties from the
uncorrelated background estimate are at or below the 1\% level.

\begin{table}
\caption{\label{tab:uncorrsys}
The systematic uncertainties for the uncorrelated background in the
separate $\dstplnu$ and $\dstzlnu$ fits and the combined fit. The upper-vertex
contribution, the lepton normalizations, and $\chi_d$ are treated as completely
correlated between the two modes; all others are uncorrelated. }
\begin{ruledtabular}
\begin{tabular}{lccc} 
Variation  & $\vcbf$(\%)&$\rho^2$(\%) & $\Gamma(\bdstlnu)$(\%) \\ 
\hline
$\dstplnu$ fit &&&\\
\hspace{5mm}Upper-vertex $D^*$        & 0.7 & 0.9 & 0.4 \\
\hspace{5mm}Other                     & 0.2 & 0.2 & 0.3 \\\hline
\hspace{5mm}Total                     & 0.7 & 1.0 & 0.5 \\\hline\hline
$\dstzlnu$ fit &&&\\
\hspace{5mm}Upper-vertex $D^*$        & 0.6 & 0.9 & 0.4 \\
\hspace{5mm}Other                     & 0.5 & 0.8 & 0.3 \\\hline
\hspace{5mm}Total                     & 0.8 & 1.2 & 0.5 \\\hline\hline
Combined fit&&&\\
\hspace{5mm}Upper-vertex $D^*$        & 0.6 & 0.9 & 0.4  \\
\hspace{5mm}Other                     & 0.3 & 0.3 & 0.3 \\\hline
\hspace{5mm}Total                     & 0.7 & 1.0 & 0.5 \\
\end{tabular}
\end{ruledtabular}
\end{table}

\subsubsection{Correlated background}
We assess the uncertainty arising from the correlated background
by varying the branching fractions of the contributing modes simultaneously 
by 50\%. Since this is a small background, this variation has little
effect on $\vcbf$, and the uncertainties are 0.1\%, 0.6\%, and 0.8\%
on $\vcbf$, $\rho^2$, and $\Gamma(\bdstlnu)$, respectively.

\subsubsection{Fake lepton background}
We vary the measured electron and muon fake rates separately by 50\%. 
This is conservative, but it has also almost no effect on our result; the
total uncertainty on $\vcbf$ is 0.02\%, while the uncertainties on  
$\rho^2$ and $\Gamma(\bdstlnu)$ are 0.3\% and 0.2\%, respectively.

\subsection{Slow $\pi$ reconstruction uncertainty}
\label{sec:slowpi}
\begin{figure*}
\epsfig{file=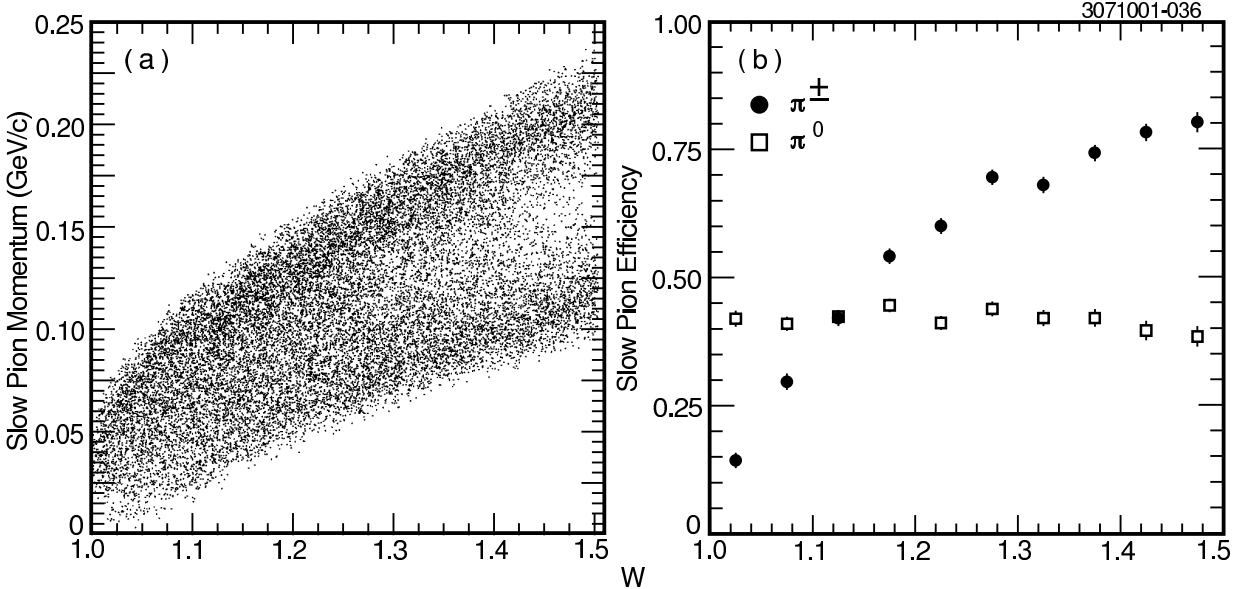,width=165mm}
\caption{\label{fig:slowpieff}
(a) The correlation between the slow pion momentum and $w$;
(b) The reconstruction efficiency as a function of $w$ for 
charged slow pions (solid circles) and neutral
slow pions (open squares).}
\end{figure*}

The largest source of uncertainty for the analysis is the efficiency
for reconstructing the slow pion from the $D^*$ decay. 
Because of the small energy release in $D^{*}$ decays, the daughter
pion has low momentum and travels approximately in the
direction of the parent $D^*$.  For our signal decays, the momentum
range of the 
slow pion is 0 to about 250~MeV/$c$. Note also that $w=E_{D^*}/m_{D*}$
in the $B$  
rest frame, so the slow-pion momentum is correlated with $w$ [see 
Fig.~\ref{fig:slowpieff}(a)]. 

% efficiency vs w
Charged and neutral slow-pion reconstruction efficiencies 
depend very differently on $w$. Charged pions with momenta less 
than 50 \mevc\ do not penetrate far enough into the tracking chamber to be 
reconstructed; the slow-pion reconstruction efficiency is therefore
low near $w=1$ and increases rapidly over the 
next few $w$ bins as the pion momentum increases.
Neutral slow pions, on the other hand, decay to two low-energy photons 
(30--230 \mev). The lowest-momentum $\pi^0$'s decay almost back-to-back, 
depositing about equal energy in the calorimeter. As the $\pi^0$ momentum 
increases, the Lorentz boost pushes some of the
photons below our minimum energy 
requirement of 30 \mev. The neutral slow-pion efficiency therefore drops 
slowly as $w$ increases. The slow-pion efficiencies for both
charged and neutral $\pi$'s from $\bdstlnu$ decays 
are shown as a function of $w$ in Fig.~\ref{fig:slowpieff}(b).

Because we rely on Monte Carlo simulation to estimate the slow-pion
efficiencies, we investigate possible differences between the
simulation and performance of the CLEO detector in order to estimate the
systematic uncertainty of slow-pion reconstruction.   
We consider the effect of nearby tracks and showers on slow-pion
reconstruction, comparing the efficiency of data and Monte Carlo-simulated
$\bbbar$ events to limit a systematic error due to a difference in the
``event environment'' in data and simulated events.
We also consider how much imperfect knowledge of detector material can
affect reconstruction efficiency through pion range-out, multiple
scattering, hadronic interaction, or photon conversions.
Finally, we vary parameters of the detector simulation for
the drift chambers and calorimeter to estimate the contribution to the
systematic uncertainty.

% event environment
Monte Carlo-simulated events may have a different number of drift
chamber hits or calorimeter showers than the data.  The detector activity
(track fragments or showers) near a candidate slow pion can affect the
reconstruction efficiency.  To evaluate the impact of
these ``environment effects'' on the slow-pion reconstruction efficiency,
we insert Monte Carlo-generated
slow-pion tracks or showers with kinematic distributions appropriate to
$\dstlnu$ decay into samples of hadronic events selected from our data
and from simulated $\bbbar$ events.  In each $w$ bin, we
compare the reconstruction efficiency for the tracks embedded
into data and simulated events.   For $\dstplnu$, the efficiency difference
is small, $\Delta \epsilon/\epsilon=(0.2\pm1.1)$\% integrated over $w$.
Likewise, the effect of event environment is small for $\dstzlnu$, where
we find a net efficiency difference 
$\Delta \epsilon/\epsilon=(-0.6\pm 1.1)$\%.  The uncertainties here
are from the statistics of the data and Monte Carlo comparison.
We measure the impact of the event environment by using the measured
data-Monte Carlo efficiency difference in each $w$ bin to 
modify the efficiency matrix in Eq.~\ref{eq:chi2} and repeating the fit. 
The slow-pion efficiency may depend on the track or shower
multiplicity, which is increased by one or two, respectively, by the
embedding study; we find no statistically significant evidence of this
in our studies, but we include a small uncertainty [0.3\% on $\vcbf$]
to cover this effect. 

% material
To estimate the uncertainty due to our imperfect knowledge of the
detector material inside the outer boundary of the tracking chambers,
we vary the material description of the detector by 10\% in our
simulation and remeasure the slow-pion efficiencies. This 10\%
variation of material is
based on a study that compared the polar
angle distribution of $e^+e^-\to\gamma\gamma$ events in data and simulation.
We then repeat the $\vcb$ fit using these new
efficiencies and take the excursions of $\vcbf$, $\rho^2$, and 
$\Gamma(\bdstlnu)$ as the uncertainty.

% pi+ simulation
In a similar way, we estimate the uncertainty due to our tracking
chamber and crystal calorimeter simulation.  For charged slow pions,
performance of the tracking devices is essential.  Differences in
hit efficiency and single-hit resolution between data and Monte Carlo
simulation can result in a difference in measured efficiency.  The
tracking simulation parameters are tuned using an independent sample
of charged tracks.  We vary the tracking chamber hit resolutions by
amounts determined from residual distributions in these data, and we
vary hit efficiencies according to observed differences in the data
and simulated hit efficiencies.  

% pi0 simulation
For neutral slow pions, performance of the calorimeter is important.
Here we consider differences in the $\mgg$ and 
transverse shower profile distributions used for $\pi^0$
reconstruction.  We calibrate the calorimeter energy scale at high
energy (1--5 GeV) using showers from QED event samples ($e^+e^-\to
e^+e^-$, $e^+e^-\to \gamma\gamma$, and $e^+e^- \to e^+ e^- \gamma$).
We check this scale with a sample of $\pi^0\to\gamma\gamma$ and
$\eta\to\gamma\gamma$ candidates, which should peak at the known
$\pi^0$ and $\eta$ masses.  For low-energy showers there can be
residual gain mismatches from nonlinearities and noise.  Accordingly,
we adjust the calorimeter noise and dispersion of crystal gains in the
simulation so that it reproduces the transverse spatial distributions
of the photon showers and the $\mgg$ distribution for an independent
sample of low-momentum $\pi^0$'s.  We vary the noise and gain dispersion
parameters in the simulation within a range determined from the data to
assess the systematic uncertainty.
% outer material
Photons that convert and begin to shower just in front of the
calorimeter will have degraded resolution.  We vary the material
description between the outer tracking chamber boundary and the
calorimeter crystals by a conservative 15\% to determine its
contribution to the uncertainty for slow-$\pi^0$ reconstruction.
% energy threshold in geant
The transverse spatial extent of photon showers varies with the
low-energy cutoff in the shower simulation. To assess the uncertainty
from the cutoff in our simulation, we lower the minimum energy for
photon simulation by a factor of 10, from 1 \mev\ to 100 keV.

%vertexing
Finally, we assess the systematic uncertainty due to requiring a
$\dstp$ and $D^0$ vertex in the $\dstplnu$ analysis by performing the
analysis without vertexing.  In the separate $\dstplnu$ fit without
vertexing, the result for $\vcbf$ shifts by $(2.0\pm1.8)$\%, where the
uncertainty takes into account correlations between analyses with and
without vertexing.  We take the quadrature sum of the shift and its
uncertainty as a systematic error. 

We find that the largest contributions to the uncertainty on $\vcbf$ come
from the material description (1.3\%), the effects of vertexing
(1.5\%), and the minimum energy for photon simulation (0.6\%).  The
statistical uncertainty from data and Monte Carlo comparisons also
contributes (1.3\%).  The given uncertainties apply to the combined fit.
Table~\ref{tab:slowpi} summarizes the uncertainties on slow-pion
reconstruction.  

\begin{table*}
\caption{\label{tab:slowpi}
The systematic errors from the slow-pion reconstruction efficiency
for the separate $\dstplnu$ and $\dstzlnu$ fits and the combined fit. We take 
the uncertainty from the material description to be correlated between the
two modes; all other errors are uncorrelated.
}
\begin{ruledtabular}
\begin{tabular}{lccc}
Mode  & $\vcbf$(\%)&$\rho^2$(\%) & $\Gamma(\bdstlnu)$(\%) \\ 
\hline
$\dstplnu$ fit&&&\\
\hspace{5mm} Material description            & 2.6 & 3.2 & 2.1 \\ 
\hspace{5mm} Tracking chamber hit efficiency & 0.6 & 0.2 & 1.4 \\ 
\hspace{5mm} Vertexing                       & 2.7 & 2.7 & 2.9 \\ 
\hspace{5mm} Other uncertainties             & 0.8 & 1.1 & 0.7 \\ 
\hspace{5mm} Statistics (environment)        & 1.7 & 2.5 & 1.4 \\\hline
\hspace{5mm} Total                           & 4.2 & 5.0 & 4.1 \\\hline\hline
$\dstzlnu$ fit&&&\\
\hspace{5mm} Material description     & 1.1 & 3.0 & 0.6 \\
\hspace{5mm} Photon cutoff            & 1.5 & 0.9 & 2.3 \\
\hspace{5mm} Other uncertainties      & 1.2 & 2.9 & 5.0 \\ 
\hspace{5mm} Statistics (environment) & 2.1 & 3.3 & 2.7 \\\hline
\hspace{5mm} Total                    & 3.1 & 5.4 & 6.2 \\ \hline\hline
Combined fit& & &\\
\hspace{5mm} Material description                          & 1.3 & 1.5 & 1.2 \\
\hspace{5mm} Tracking chamber hit efficiency ($\dstplnu$ only)& 0.3&0.2 & 0.9\\
\hspace{5mm} Vertexing ($\dstplnu$ only)                   & 1.5 & 1.6 & 1.7\\
\hspace{5mm} Photon cutoff ($\dstzlnu$ only)               & 0.6 & 0.2 & 0.9 \\
\hspace{5mm} Other uncertainties                           & 0.9 & 1.0 & 1.8 \\
\hspace{5mm} Statistics (environment)                      & 1.3 & 2.1 & 1.3 \\
\hline
\hspace{5mm} Total                                         & 2.6 & 3.1 & 3.3 \\
\end{tabular}
\end{ruledtabular}
\end{table*}

\subsection{Sensitivity to $R_1(1)$ and $R_2(1)$}
The form factor ratios $R_1(1)$ and $R_2(1)$ affect the lepton spectrum 
and therefore the fraction of decays satisfying our 0.8~GeV/$c$ electron and 
1.4~GeV/$c$ muon momentum requirements. They also affect the relative 
contributions of the three $\dstlnu$ form factors, and therefore can affect 
the form-factor slope $\rho^2$.

To estimate the uncertainty due to the measurement errors on $R_1(1)$ and 
$R_2(1)$, we use
\begin{equation}
\sigma_{P}^2 = \sum_{i,j=1}^2 \frac{\partial P}{\partial R_i(1)}
\frac{\partial P}{\partial R_j(1)} E_{ij},
\end{equation}
where $P$ stands for the parameter 
[$\vcbf$, $\rho^2$, or $\Gamma (\bdstlnu)$]
whose uncertainty we are calculating,
$E_{ii} = \sigma_i^2$ and $E_{ij} = \rho_{ij} \sigma_i \sigma_j$, where 
$\rho_{12}=-0.82$ is the correlation coefficient from the 
$R_1(1)$ and $R_2(1)$ measurement~\cite{ffprl}. We compute the
partial derivatives
$\partial P/\partial R_i(1)$ by shifting $R_i$ and repeating our analysis.
We find an uncertainty on $\vcbf$ from
this source of 1.4\%, and a substantial uncertainty on $\rho^2$ of 12\%.

\subsection{Other uncertainties}
We considered the following minor sources of systematic uncertainty,
summarized in Table~\ref{tab:syserr_all}.

The efficiency for identifying electrons has been evaluated using radiative 
Bhabha events embedded in hadronic events, and has an uncertainty of 2.6\%. 
Similarly, the muon identification efficiency has been evaluated using 
radiative mu-pair events, and has an uncertainty of 1.6\%.  We
determine the total uncertainty from lepton identification by adding in
quadrature the shift in results when repeating the analysis with
electron and muon efficiencies varied by their momentum-dependent
uncertainties. Separate electron and muon analyses of our data give 
the results shown in Table~\ref{tab:e_mu_fits}.
Including the systematic uncertainties on lepton identification, the
separate electron and muon results are consistent at the 35\%
confidence level.

\begin{table}
\caption{\label{tab:e_mu_fits}
The results from separate analyses using only $D^* e^-\bar{\nu}$ or
$D^* \mu^-\bar{\nu}$. The errors are statistical only. In these fits
$f_{+-}$ is constrained using Ref.~\protect\onlinecite{sylvia}.}
\begin{ruledtabular}
\begin{tabular}{lccc} 
Mode & $\vcbf$(\%)&$\rho^2$(\%) & $\Gamma(\bdstlnu)$(\%) \\ 
\hline
$\dstplnu$ &&&\\
\hspace{5mm}$D^{*+}  e^-\bar{\nu}$ & 0.0420$\pm$0.0023 & 1.65$\pm$0.14 & 0.0363$\pm$0.0021\\
\hspace{5mm}$D^{*+}\mu^-\bar{\nu}$ & 0.0448$\pm$0.0026 & 1.69$\pm$0.15 & 0.0404$\pm$0.0025\\\hline
$\dstzlnu$ &&&\\
\hspace{5mm}$D^{*0}  e^-\bar{\nu}$ & 0.0409$\pm$0.0032 & 1.41$\pm$0.24 & 0.0396$\pm$0.0030\\
\hspace{5mm}$D^{*0}\mu^-\bar{\nu}$ & 0.0474$\pm$0.0040 & 1.80$\pm$0.26 & 0.0423$\pm$0.0042\\\hline
combined &&&\\
\hspace{5mm}$D^{*}  e^-\bar{\nu}$ & 0.0420$\pm$0.0018 & 1.60$\pm$0.12 & 0.0374$\pm$0.0015\\
\hspace{5mm}$D^{*}\mu^-\bar{\nu}$ & 0.0457$\pm$0.0021 & 1.73$\pm$0.13 & 0.0411$\pm$0.0019\\
\end{tabular}
\end{ruledtabular}
\end{table}

The $B$ momentum is measured directly in the data using fully reconstructed 
hadronic decays, and is known on average with a precision of 0.0016~GeV/$c$.
Variation of the momentum in our reconstruction slightly alters the $\cby$ 
distribution that we expect for our signal, and it therefore changes the 
yields obtained from the $\cby$ fits. Likewise, CLEO has measured the $B^0$-
and $B^+$-meson masses~\cite{ershov} and when we vary them within their
measurement errors, we find a small effect on the yields.

We determine the tracking efficiency uncertainties for the lepton and the 
$K$ and $\pi$ forming the $D^0$ in the same study used for the slow pion
from the $\dstp$ decay.  These uncertainties are confirmed in a study of 
1-prong versus 3-prong $\tau$ decay events from our data sample.

The final-state radiation model has a small effect on our $\dstlnu$ yields 
because it affects the $\dstlnu$ $\cby$ distributions. 
Because we require $p_e \ge 0.8$ GeV/$c$ and $p_\mu \ge 1.4$ GeV/$c$, 
the model also affects the $\dstlnu$ efficiency. The final-state
radiation model is estimated by the authors of \textsc{Photos} to be accurate
within 30\%~\cite{photos}. We determine our 
sensitivity to the model by repeating our analysis without including 
radiative $\dstlnu$ decays in our $\dstlnu$ Monte Carlo. We then take
30\% of the change to our results as our uncertainty. 

\begin{figure}
\epsfig{file=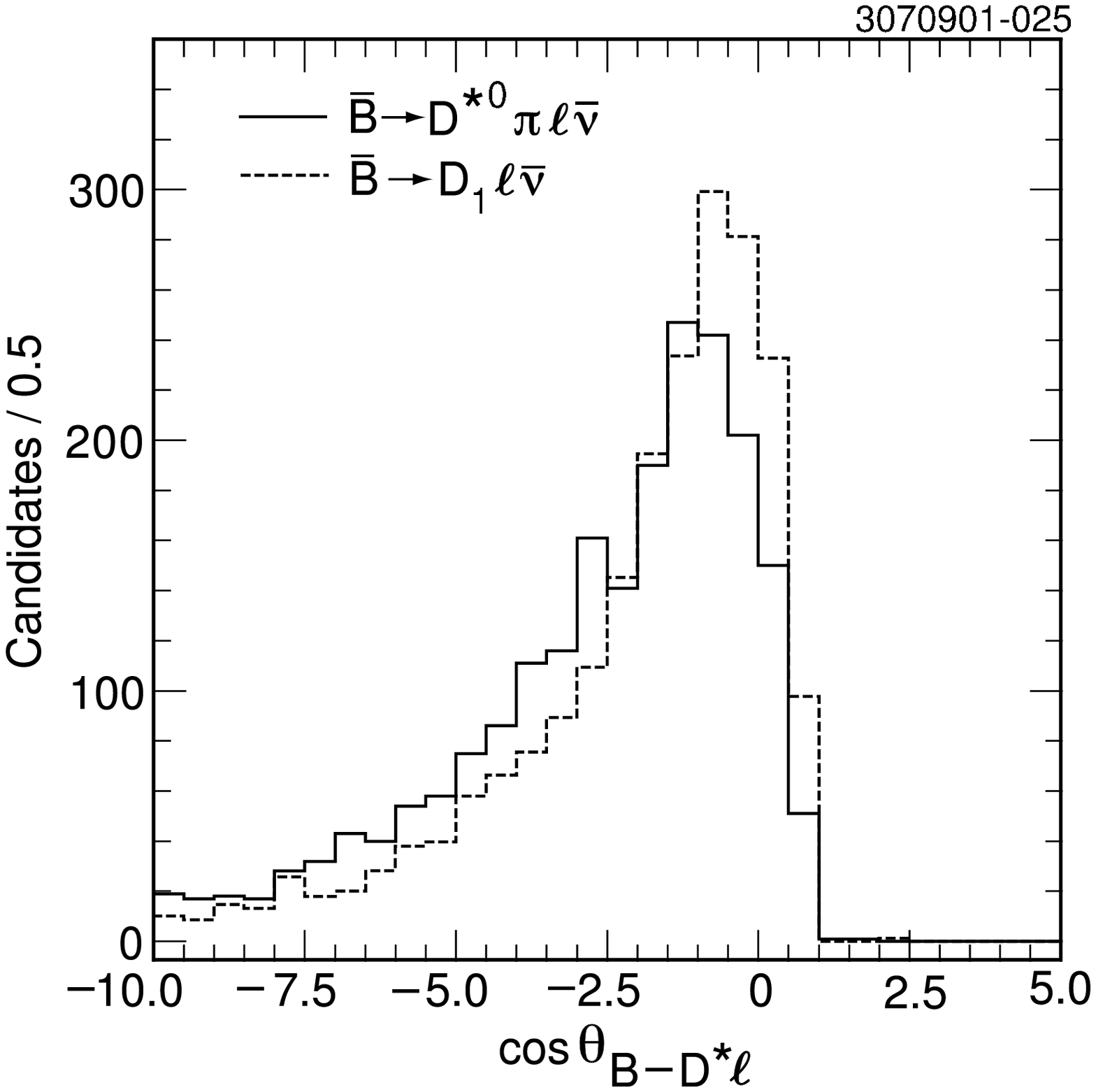,width=3.4in}
\caption{\label{fig:ddstlnu}
The $\cby$ distribution of 
$\bar{B}\to D^{*0}\pi\ell\bar{\nu}$ (solid histogram) 
and $\bar{B}\to D_1\ell\bar{\nu}$ (dashed histogram) decays
contributing to the $\dstxlnu$ 
sample for $\dstzlnu$. The histograms are normalized to equal area.}
\end{figure}

Finally, our analysis requires that we know the $\cby$ distribution of the 
$\dstxlnu$ contribution.  This distribution in turn depends on both the 
branching fractions of contributing modes and on their form factors. 
Variation of all of these branching fractions and form factors is not only 
cumbersome, but also out of reach given the poor current knowledge of
these modes. 
Instead, we note that the $\bar{B}\to D^*\pi\ell\bar{\nu}$ and $\bar{B}\to D_1\ell\bar{\nu}$ modes 
are the ones with the most extreme $\cby$ distributions (the largest mean and 
the smallest). These distributions are shown in Fig.~\ref{fig:ddstlnu}.  
We therefore repeat the analysis, first using only $\bar{B}\to
D^*\pi\ell\bar{\nu}$ to describe our $\dstxlnu$ decays and then
using only $\bar{B}\to D_1\ell\bar{\nu}$ to describe these decays;
we take the larger of the two excursions as our systematic error.

\section{Conclusions}

\begin{figure}
\epsfig{file=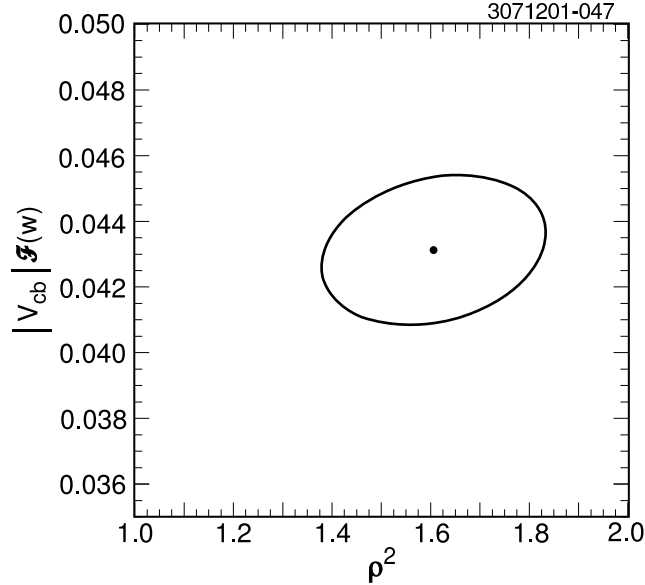,width=3.4in}
\caption{\label{fig:ellipse}
The error ellipse for the combined $\dstplnu$ and $\dstzlnu$
measurement, including statistical and systematic uncertainties.}
\end{figure}

We have fit the $w$ distribution of $\bdstlnu$ decays for 
the slope of the form factor and $\vcbf$.
For the combined $\dstplnu$ and $\dstzlnu$ fit, we find
\begin{eqnarray}
\vcbf &=& 0.0431 \pm 0.0013 \pm 0.0018,\ {\text{and}} \nonumber\\
\rho^2       &=& 1.61   \pm 0.09   \pm 0.21.  \nonumber
\end{eqnarray}
Including the systematic uncertainties we compute a
correlation coefficient $C(\vcbf, \rho^2)=0.22$.
Figure~\ref{fig:ellipse} shows the total error ellipse for this measurement.
The best-fit parameters imply the decay rate
\begin{equation}
\Gamma = 0.0394\pm 0.0012\pm 0.0026 \ \text{ps}^{-1}. \nonumber
\end{equation}

We recover the branching fractions from the rate by dividing
by the appropriate $B$-meson lifetimes.  These results are
sensitive only to the ratio of $B^+$ to $B^0$ lifetimes.
They are
\begin{eqnarray}
{\cal B}(\bdstplnu) &=& (6.09 \pm 0.19 \pm0.40)\% {\ \text{and}}  \nonumber\\
{\cal B}(\bdstzlnu) &=& ( 6.50 \pm 0.20 \pm0.44)\%, \nonumber
\end{eqnarray}
where the errors are completely correlated. 

A recent lattice calculation yields~\cite{kronfeld}
$\fone=0.919^{+0.030}_{-0.035}$ after applying a QED correction
of $+0.007$.  This value is consistent with 
$\fone=0.913\pm 0.042$, the evaluation of the authors of
Ref.~\onlinecite{babarbook}, but is more precise.
Using the lattice value of $\fone$, our result implies
\begin{equation}
\vcb =0.0469 \pm 0.0014 \text{(stat.)} \pm 0.0020 \text{(syst.)}\pm
0.0018 \text{(theo.)}. \nonumber
\end{equation}
Since full radiative corrections have yet to be calculated for
$\fone$, it is ambiguous how best to treat radiative decays in
the analysis.  We include radiative decays in our signal.

This value of $\vcb$ is consistent 
with previous values obtained from $\dstlnu$ decays
\cite{aleph_dslnu,opal_dslnu,delphi_dslnu,belle_dslnu}, but is
somewhat higher.
However, we note our ability to reconstruct $\cby$ makes our
analysis approximately four times less sensitive to the poorly known
$\dstxlnu$ background, and furthermore allows us to constrain it
with the data.
This value of $\vcb$ is also somewhat higher than that obtained
using inclusive semileptonic $B$ decays~\cite{cleomoments}.  If
confirmed, 
this discrepancy could signal a violation of quark-hadron duality.
A larger value of $\vcb$ shifts constraints on the CKM unitarity
triangle from $|V_{ub}/V_{cb}|$ and $CP$ violation in the neutral kaon
system, somewhat reducing expectations for indirect $CP$ violation
in the $B$ system.
  
\begin{acknowledgments}
We gratefully acknowledge the effort of the CESR staff in providing us with
excellent luminosity and running conditions.
We thank A.~Kronfeld, W.~Marciano, and M.~Neubert for helpful discussions.
M.~Selen thanks the PFF program of the NSF and the Research Corporation, 
and A.~H.~Mahmood thanks the Texas Advanced Research Program.
This work was supported by the National Science Foundation and the
U.~S.~Department of Energy.
\end{acknowledgments}


\begin{thebibliography}{99}

\bibitem{Ckm} 
N.~Cabibbo, \PRL{10}{531}{1963}.
%%CITATION = PRLTA,10,531;%%

\bibitem{cKM}
M.~Kobayashi and T.~Maskawa, Prog.\ Theor.\ Phys. \textbf{49}, 652 (1973).
%%CITATION = PTPKA,49,652;%%

\bibitem{unitaritytriangle} 
L.~-L.~Chau and W.~-Y.~Keung, \PRL{53}{1802}{1984}; 
%%CITATION = PRLTA,53,1802;%%
J.~D.~Bjorken, \PRD{39}{1396}{1989}; 
%%CITATION = PHRVA,D39,1396;%%
C.~Jarlskog and R.~Stora, \PLB{208}{268}{1988}; 
%%CITATION = PHLTA,B208,268;%%
J.~L.~Rosner, A.~I.~Sanda, and M.~P.~Schmidt, in 
\textit{Proceedings of the Workshop on High Sensitivity Beauty Physics at Fermilab}. Fermilab, November 11-14, 1987, edited by A.~J.~Slaughter, N.~Lockyer, and M.~Schmidt (Fermilab, Batavia, 1988), p 165; 
C.~Hamzaoui, J.~L.~Rosner, and A.~I.~Sanda, \textit{ibid.}, p 215.

\bibitem{K0cpv}
See, for example, A.~J.~Buras, M.~E.~Lautenbacher, and G.~Ostermaier,
\PRD{50}{3433}{1994}.
%%CITATION = HEP-PH 9403384;%%

\bibitem{hqs_voloshin} % considers heavy quark symmetry, 
		       % including semileptonic decays
M.~B.~Voloshin and M.~A.~Shifman, Sov. J. Nucl. Phys. {\textbf 47}, 511 (1988).
%%CITATION = SJNCA,47,511;%%

\bibitem{hqs_isgur} % applies heavy quark symmetry to weak decays
		    % (Isgur Wise function)
N.~Isgur and M.~B.~Wise, \PLB{232}{113}{1989}; {\textbf 237}, 527 (1990).
%%CITATION = PHLTA,B232,113;%%
%%CITATION = PHLTA,B237,527;%%

\bibitem{hqs_luke} % Luke's theorem; applied hqs to 1/Mq expansion;
		   % Considers D* l nu and zero-recoil, Vcb
M.~E.~Luke, \PLB{252}{447}{1990}.
%%CITATION = PHLTA,B252,447;%%

\bibitem{hqs_falk} % heavy to heavy semileptonic decays HQS
A.~F.~Falk, H.~Georgi, B.~Grinstein, and M.~B.~Wise, \NPB{343}{1}{1990}.
%%CITATION = NUPHA,B343,1;%%

\bibitem{hqs_neubert} % considers Vcb from D*lnu
M.~Neubert, \PLB{264}{455}{1991}. %\textit{ibid.} {\textbf 264}, 455 (1991).
%%CITATION = PHLTA,B264,455;%%

%\bibitem{hqet} A.V.Manohar and M.B.Wise, \textit{Heavy Quark Physics}
%(Cambridge University, Cambridge, 2000) Chap. 2.

\bibitem{hqet_grinstein} % HQET formulation as effective field theory
B.~Grinstein, \NPB{339}{253}{1990}.
%%CITATION = NUPHA,B339,253;%%

\bibitem{hqet_eichten} % HQET formulation as effective field theory
E.~Eichten and F.~Feinberg, \PRD{23}{2724}{1981};
%%CITATION = PHLTA,B264,455;%%
E.~Eichten and B.~Hill, \PLB{234}{511}{1990}; \textbf{243}, 427 (1990).
%%CITATION = PHLTA,B234,511;%%
%%CITATION = PHLTA,B243,427;%%

\bibitem{hqet_georgi} % HQET formulation as effective field theory
H.~Georgi, \PLB{240}{447}{1990}.
%%CITATION = PHLTA,B240,447;%%

\bibitem{hqet_koerner} % HQET formulation as effective field theory
J.~G.~K\"{o}rner and G.~Thompson, \PLB{264}{185}{1991}.
%%CITATION = PHLTA,B264,185;%%

\bibitem{hqet_mannel} % HQET formulation as effective field theory
T.~Mannel, W.~Roberts, and Z.~Ryzak, \NPB{368}{204}{1992}.
%%CITATION = NUPHA,B368,204;%%

\bibitem{dslnu_neubert} % HQET applied to D* l nu
                        % perturbative QCD HQET matching for F(1)
M.~Neubert, \PRD{46}{2212}{1992}.
%%CITATION = PHRVA,D46,2212;%%

\bibitem{dslnu_falk}  % HQET applied to D* l nu
                      % power corrections 1/mQ^2 for D(*) l nu
A.~F.~Falk and M.~Neubert, \PRD{47}{2965}{1993}.
%%CITATION = HEP-PH 9209268;%%

\bibitem{letter} 
CLEO Collaboration, R.~A.~Briere \etal, \PRL{89}{081803}{2002}
[hep-ex/0202032].
%%CITATION = HEP-EX 0203032;%%

\bibitem{oldcleo}
CLEO Collaboration, B.~Barish \etal,  \PRD{51}{1014}{1995}.
%%CITATION = HEP-EX 9406005;%%

\bibitem{blvthesis}
B.~L.~Valant-Spaight, Ph.~D.~thesis, Cornell University, 2001.

\bibitem{cleonim} 
CLEO Collaboration, Y.~Kubota \etal, Nucl. Instrum. Methods
Phys.  Res., Sect. A \textbf{320}, 66 (1992).
%%CITATION = NUIMA,A320,66;%%

\bibitem{geant} 
R.~Brun \textit{et al.}, \textsc{Geant} 3.15, CERN DD/EE/84-1.

\bibitem{r2}
G.~C.~Fox and S.~Wolfram, \PRL{41}{1581}{1978}.
%%CITATION = PRLTA,41,1581;%%

\bibitem{Kalman} 
P.~Billoir, Nucl. Inst. Meth. Res., Sect. A \textbf{255}, 352 (1984).
%%CITATION = NUIMA,A225,352;%%

\bibitem{ershov}
CLEO Collaboration, S.E.~Csorna \etal, \PRD{61}{111101}{2000} [hep-ex/0001013].
%%CITATION = HEP-EX 0001013;%%

\bibitem{likelihood}
R.~Barlow and C.~Beeston, Comput. Phys. Commun. \textbf{77}, 219 (1993).
%%CITATION = CPHCB,77,219;%%

\bibitem{caprini}
I.~Caprini, L.~Lellouch, and M.~Neubert, 
Nucl. Phys. \textbf{B530}, 153 (1998) [hep-ph/9712417]. 
%%CITATION = HEP-PH 9712417;%%

\bibitem{photos}
E.~Barberio and Z.~W\c{a}s, Comput. Phys. Commun. \textbf{79}, 291 (1994).
%%CITATION = CPHCB,79,291;%%

\bibitem{isgw2}
D.~Scora and N.~Isgur, \PRD{52}{2783}{1995};
%%CITATION = HEP-PH 9503486;%%
N.~Isgur \etal, \PRD{39}{799}{1989}.
%%CITATION = PHRVA,D39,799;%%

\bibitem{goityroberts}
J.~L.~Goity and W.~Roberts, \PRD{51}{3459}{1995}.
%%CITATION = HEP-PH 9406236;%%

\bibitem{pdg}
D.~E.~Groom \etal, Eur. Phys. J. C \textbf{15}, 1 (2000).
%%CITATION = EPHJA,C15,1;%%

\bibitem{DDK}
CLEO Collaboration, CONF 97-26 (1997).

\bibitem{roywang}
CLEO Collaboration, B.~Barish \etal, \PRL{76}{1570}{1996}.
%%CITATION = PRLTA,76,1570;%%

\bibitem{alephdstx}
ALEPH Collaboration, D.~Buskulic \etal, \ZPC{73}{601}{1997}.
%%CITATION = ZEPYA,C73,601;%%

\bibitem{delphidstx}
DELPHI Collaboration, P.~Abreu \etal, \PLB{475}{407}{2000}.
%%CITATION = HEP-EX 0105052;%%

\bibitem{richburch}
J.~D.~Richman and P.~R.~Burchat, Rev. Mod. Phys. \textbf{67}, 893 (1995).
%%CITATION = HEP-PH 9508250;%%

\bibitem{ffprl}
CLEO Collaboration, J.~Duboscq \etal, \PRL{76}{3898}{1996}. 
%%CITATION = PRLTA,76,3898;%%

\bibitem{neubert}
M.~Neubert, Physics Reports, \textbf{245}, 259 (1994).
%%CITATION = HEP-PH 9306320;%%

\bibitem{lebed}
C.~G.~Boyd, B.~Grinstein, and R.~F.~Lebed, \PRD{56}{6895}{1997} [hep-ph/9705252].
%%CITATION = HEP-PH 9705252;%%

\bibitem{cleo_kpi}
CLEO Collaboration, D.~S.~Akerib \etal, \PRL{71}{3070}{1993}.
%%CITATION = PRLTA,71,3070;%%

\bibitem{aleph_kpi}
ALEPH Collaboration, R.~Barate \etal, \PLB{403}{367}{1997}.
%%CITATION = PHLTA,B403,367;%%

\bibitem{sylvia}
CLEO Collaboration, J.~P.~Alexander \etal, \PRL{86}{2737}{2001},
[hep-ex/0006002].
%%CITATION = HEP-EX 0006002;%%

\bibitem{kronfeld}
S. Hashimoto \etal, \PRD{66}{014503}{2002}, [hep-ph/0110253].
%%CITATION = HEP-PH 0110253;%%

\bibitem{babarbook}
BaBar Physics Book, edited by P.~F.~Harrison and H.~R.~Quinn,
SLAC-R-504 (1998).

\bibitem{aleph_dslnu}
ALEPH Collaboration, D.~Buskulic \etal, \PLB{395}{373}{1997}.
%%CITATION = PHLTA,B395,373;%%

\bibitem{opal_dslnu}
OPAL Collaboration, G.~Abbiendi \etal, \PLB{482}{15}{2000}.
%%CITATION = HEP-EX 0003013;%%

\bibitem{delphi_dslnu}
DELPHI Collaboration, P.~Abreu \etal, \PLB{510}{55}{2001}.
%%CITATION = HEP-EX 0104026;%%

\bibitem{belle_dslnu}
BELLE Collaboration, K.~Abe \etal, \PLB{526}{258}{2002}.
%%CITATION = HEP-EX 0111082;%%

\bibitem{cleomoments}
CLEO Collaboration, D.~Cronin-Hennessy \etal, \PRL{87}{251808}{2001},
[hep-ex/0108033].
%%CITATION = HEP-EX 0108033;%%

\end{thebibliography}
\end{document}